\documentclass[submission]{SciPost}

\usepackage{amsfonts}
\usepackage{mathrsfs}
\usepackage{tikz}
\usepackage{slashed}
\usepackage{caption}
\usepackage{subcaption}

\newcommand{\eq}[1]{\begin{align}#1\end{align}}

\def\m{\mu}

\def\d{\partial}

\usetikzlibrary{arrows.meta, positioning, quotes}
\usetikzlibrary{arrows}
\usetikzlibrary{decorations.markings,intersections}
\usetikzlibrary{decorations.pathmorphing}
\usetikzlibrary{shapes.misc}
\usetikzlibrary{snakes}
\tikzset{cross/.style={cross out, draw=black, minimum size=2*(#1-\pgflinewidth), inner sep=0pt, outer sep=0pt},
cross/.default={1pt}}
\usepackage{tikz-cd}

\begin{document}

\begin{center}{\Large \textbf{
Fractons and exotic symmetries from branes 
}}\end{center}

\begin{center}
Hao Geng\textsuperscript{1,2},
Shamit Kachru\textsuperscript{3},
Andreas Karch\textsuperscript{2,4},
Richard Nally\textsuperscript{3},
Brandon C. Rayhaun\textsuperscript{3}
\end{center}

\begin{center}
{\bf 1} Harvard University
\\
{\bf 2} University of Washington
\\
{\bf 3} Stanford University
\\
{\bf 4} University of Texas at Austin
\\
\end{center}

\begin{center}
\today
\end{center}

\section*{Abstract}
{\bf The emerging study of fractons, a new type of quasi-particle with restricted mobility, has motivated the construction of several classes of interesting continuum quantum field theories with novel properties. One such class consists of \emph{foliated field theories} which, roughly, are built by coupling together fields supported on the leaves of foliations of spacetime. Another approach, which we refer to as \emph{exotic field theory}, focuses on constructing Lagrangians consistent with special symmetries (like subsystem symmetries) that are adjacent to fracton physics. A third framework is that of \emph{infinite-component Chern-Simons theories}, which attempts to generalize the role of conventional Chern-Simons theory in describing (2+1)D Abelian topological order to fractonic order in (3+1)D. The study of these theories is ongoing, and many of their properties remain to be understood.

Historically, it has been fruitful to study QFTs by embedding them into string theory. One way this can be done is via D-branes, extended objects whose dynamics can, at low energies, be described in terms of conventional quantum field theory. QFTs that can be realized in this way can then be analyzed using the rich mathematical and physical structure of  string theory. In this paper, we show that foliated field theories, exotic field theories, and infinite-component Chern-Simons theories can all be realized on the world-volumes of branes. We hope that these constructions will ultimately yield valuable insights into the physics of these interesting field theories.
}

\vspace{10pt}
\noindent\rule{\textwidth}{1pt}
\tableofcontents\thispagestyle{fancy}
\noindent\rule{\textwidth}{1pt}
\vspace{10pt}

\section{Introduction}
\label{sec:intro}

There is a long history of using string theory to gain insights into the physics of quantum field theory (see e.g.\ \cite{giveon1999brane} for a review). Central to this program is the notion of a D$p$-brane \cite{polchinski1995dirichlet}, an extended object with $(p+1)$ spacetime dimensions on which the endpoint of an open string can anchor. At long distances, strings resemble point particles, and so their interactions are described by a conventional low energy effective field theory which lives on the world-volume of the D-brane (or the intersections of the D-brane\emph{s}) to which they are attached. If one can realize a field theory on a configuration of branes in this manner, one can leverage the rich structure of string theory to reason about its physics.

One classic example of this philosophy in action is the work of Hanany--Witten \cite{hanany1997type}, who showed (among other things) that mirror symmetry of (2+1)D $\mathcal{N}=4$ supersymmetric gauge theories \cite{intriligator1996mirror} is a consequence of the $\textsl{SL}(2,\mathbb{Z})$ S-duality of Type IIB string theory, and were even able to predict new examples. Non-supersymmetric versions of these dualities arise in condensed matter physics \cite{seiberg2016duality,Karch:2016sxi}, and can even be deduced in some cases from mirror symmetry by adding supersymmetry breaking perturbations to a supersymmetric parent duality \cite{tong2000dynamics,kachru2016bosonization,kachru2017nonsupersymmetric}.

Recently, there has been significant energy invested in understanding the physics of a new kind of emergent quasiparticle: the so-called ``fracton'' (see e.g.\ \cite{review1,review2} for recent reviews). First discovered in the context of quantum lattice models \cite{chamon,haah1,haah2,haah3}, fracton theories often have several novel and exotic properties which make them challenging to study using continuum methods. One example is that the motion of fractons is restricted to be along rigid, positive codimension submanifolds of space. Another is that the ground state degeneracy of fracton theories usually grows exponentially with the linear system size in a way which diverges in the continuum limit. A third property is that they enjoy global, codimension-$k$ subsystem symmetries: these are symmetries which are implemented by operators supported on rigid codimension-$k$ submanifolds of space, in contrast with ordinary symmetries whose corresponding operators are defined on all of space. While many models exhibit all of these properties together, this is not always the case. Of particular interest for us will be the phenomenon of subsystem symmetries, irrespective of whether or not it is accompanied by limited mobility excitations. The goal of our paper then is to offer paradigms for realizing these kinds of unconventional theories on the world-volumes of branes, and in so doing, continue in the rich tradition of mining string theory for intuitions about field theory. 

In light of the various obstacles highlighted in the previous paragraph, there have been several novel continuum frameworks proposed for describing fracton physics. In this paper, we focus on the following three approaches.

\paragraph{Foliated field theories}
The starting point for the first approach is the recognition that it is possible to obtain a $(d+1)$-dimensional theory with a codimension-$n$ subsystem symmetry group $G$ by foliating space with $(d-n+1)$-dimensional theories, each having an \emph{ordinary} symmetry $G$.  Such a theory is somewhat trivial: it is simply a collection of decoupled lower dimensional theories. However one can produce a more interesting phase by immersing these decoupled constituents in a conventional bulk $(d+1)$-dimensional theory which mediates interactions between them. This bulk theory is often a gauge theory which imposes some kind of relation between the symmetries on the different foliations. A continuum limit can then be taken so that the spacing between the layers goes to zero; this results in non-standard ``foliated fields'', and the field theory so-produced is referred to as a \emph{foliated field theory} \cite{slagle2019foliated,slaglefoliated2,hsin2021comments} (see also \cite{aasen2020topological}).

\paragraph{Exotic field theories}

Another approach, elucidated in a series of recent papers \cite{seibergshao1,seibergshao2,seibergshao3,seibergshao4,seibergshao5,seibergshao6,seibergshao7}, involves thinking only about standard $(d+1)$-dimensional fields, but imposing non-standard symmetries on them, like subsystem symmetries. To construct Lagrangians, one proceeds in the usual way by enumerating all relevant and marginal terms consistent with the symmetry, which typically requires sacrificing relativistic invariance, and often even rotational invariance. We will call such field theories \emph{exotic field theories}, borrowing the adjective used to describe their symmetries in \cite{seibergshao1,seibergshao2,seibergshao3,seibergshao4}.

\paragraph{Infinite-component Chern-Simons theories} The last approach \cite{shirley2020twisted,ma2020fractonic} attempts to generalize the success of Abelian Chern--Simons theory in describing Abelian topological order in (2+1)D. In particular, the authors consider an infinite number of $\textsl{U}(1)$ gauge fields $a^I$, coupled together by Chern--Simons terms, and  interpret the resulting theory as giving rise to fractonic order in (3+1)D by thinking of the index $I$ as an emergent spatial dimension. We follow the authors in referring to such theories as \emph{infinite--component Chern-Simons theories}, or iCS theories for short. \\

In this paper, for each of the three approaches outlined above, we offer a few examples of intersecting brane configurations whose world-volumes furnish field theories of the corresponding type. It is our hope that these models can form the basis for future investigations into this subject.

The structure of the rest of this article is as follows. In \S\ref{sec:branereminders}, we offer a telegraphic review of the basics of D-branes in string theory. Section \S\ref{sec:fft} is devoted to foliated field theories, with a focus on linear and planar $\textsl{U}(1)$ subsystem symmetries in (2+1)D and (3+1)D, respectively. In both cases, we warm up by reviewing non-supersymmetric toy models, and then construct brane models which realize supersymmetric versions of these toy models on their world-volumes. In \S\ref{sec:eft}, we study a class of exotic field theories in (2+1)D with subsystem symmetries supported at every point in space. Again, we start by establishing intuitions on non-supersymmetric prototypes, before moving on to a supersymmetric brane web. In \S\ref{sec:ics}, we offer a brane construction of certain infinite-component Chern--Simons theories with quasi-diagonal K-matrices; interestingly, here our brane construction is able to realize the non-supersymmetric theories on the nose. Finally, we conclude in \S\ref{sec:conclusion} with suggestions for future research. 
\\

\noindent \emph{While this work was in its final stages of completion, we became aware of a recently posted paper \cite{Razamat:2021jkx} which shares some conceptual similarities with a few of our results, primarily those of \S\ref{sec:ics}.}

\section{Branes and their dynamics}\label{sec:branereminders}

As we hope to make this paper accessible to condensed matter theorists and high energy theorists alike, we start with a quick review of the basics of D$p$-branes. For more thorough reviews, see e.g.\  \cite{giveon1999brane,polchinski1996tasi,johnson2001d}.

In string theory, one often starts by positing a (1+1)D ``worldsheet quantum field theory.'' This is an action which governs embeddings of the string worldsheet into spacetime. In the simplest setting of the bosonic string, one takes this worldsheet theory to be the Polyakov action, given in the conventions of \cite{tong:strings} as  
\eq{S_{P} = -\frac{1}{4\pi\alpha'}\int d^2\sigma \sqrt{-h}h^{\alpha\beta}\d_\alpha X^\m \d_\beta X^\nu G_{\mu\nu},}
where $\alpha'=l_{s}^2$ is the square of the string length, $X^\mu$ is an embedding coordinate from the two-dimensional world-sheet into the spacetime in which the string is propagating, $G_{\mu\nu}$ is the metric on this spacetime, $\sigma^\alpha$ are the coordinates along the world-sheet of the string, and $h^{\alpha\beta}$ is its metric. This action is a small variation on the Nambu-Goto action, which computes the area of the embedded worldsheet in the same way that the standard action for a relativistic point particle computes the length of the embedded world-line. In the superstring theories which we consider in this paper, this action is supplemented by fermionic degrees of freedom. When one quantizes the worldsheet theory, one finds some finite number of light particles, as well as an infinite tower of increasingly heavy string excitations. String theory can then be thought of, from one point of view, as a machine for turning a worldsheet action into an S-matrix in spacetime which describes scattering amplitudes between these various excitations. 

There are different choices of boundary conditions one can place on the worldsheet. In this paper, we work with Type II string theories, which admit both open and closed strings. In the case of open strings, typically one takes one endpoint of the string to be restricted to lie on some submanifold of spacetime, and likewise for the other endpoint. One says that there is a D-brane (or possibly multiple D-branes) located at such submanifolds, and interprets these branes as actual dynamical objects of the string theory. By convention, we label them by their spatial dimension, e.g.\ the world-volume of a D3-brane has 3+1 spacetime dimensions. 

At low energies and long distances, strings resemble point particles, and so the part of the S-matrix which describes the \emph{massless} excitations of the string is captured by a conventional low energy effective field theory. For closed strings, this is typically Einstein-Hilbert gravity supplemented by various matter and higher--form gauge fields. For open strings, this is typically a gauge theory localized on the world-volume of the D$p$-brane; this gauge theory describes the dynamics of the massless modes of the strings whose endpoints are both anchored on the D$p$-brane. In particular, the world-volume of $N$ parallel D$p$-branes can be uniformly described, for all values of $p$, as the dimensional reduction of $\mathcal{N}=1$ super Yang-Mills theory in (9+1)D with gauge group $U(N)$ down to $(p+1)$D. For example, when $p=3$, this dimensionally-reduced theory coincides with $\mathcal{N}=4$ super Yang Mills in (3+1)D. In general, the bosonic part of the D$p$-brane action includes a Lie algebra valued gauge field $A_\mu$ as well as scalars $X^I$, for $I=1,\dots,9-p$, in the adjoint representation of the gauge group, 
\begin{align}\label{eqn:bosonicSYM}
\begin{split}
    \mathcal{L}_{\mathrm{bosonic}} &\sim \frac{1}{g_{\mathrm{YM}}^2}\mathrm{Tr}\left(\frac{1}{4}F_{\mu\nu}F^{\mu\nu}+\frac{1}{l_s^4}\mathcal{D}_\mu X^I \mathcal{D}^\mu X_I   \right), \\
    (F_{\mu\nu}^a &= \partial_\mu A_\nu^a - \partial_\nu A_\mu^a +f^{abc}A_\mu^b A_\nu^c )
\end{split}
\end{align}
where in the above, $\mathcal{D}_\mu$ is a covariant derivative, $f^{abc}$ are the structure constants of the Lie algebra in some basis, $l_s$ is the string length,  and 
\begin{align}\label{eqn:gym}
    g_{\mathrm{YM}}^2 \sim g_sl_s^{p-3}
\end{align}
with $g_s$ the string coupling. The fact that the Yang Mills coupling is proportional to $g_s$ is a straightforward consequence of the fact that the kinetic terms of the gauge field come from open string interactions, and the powers of $l_s$ simply follow on dimensional grounds. 

In addition, there is a potential governing the scalars, which schematically takes the form 
\begin{align}
    V\sim \frac{1}{l_s^8g_{\mathrm{YM}}^2}\sum_{I,J} \mathrm{Tr}[X^I,X^J]^2.
\end{align}
This potential admits a moduli space of supersymmetric vacua which is isomorphic to $(\mathbb{R}^{9-p})^{N}/S_{N}$, and is coordinatized by the eigenvalues of the $X^I$. A point in this moduli space can be thought of as specifying the locations of the $N$ D$p$-branes in the $(9-p)$ transverse directions; the quotient by the symmetric group $S_{N}$ reflects the fact that the branes are indistinguishable. If the $N$ branes are separated from each other into $m$ groups, each with $N_i$ branes, then the pattern of expectation values given to the $X^I$ induce a Higgs mechanism which spontaneously breaks the gauge group as $U(N)\to \prod_{i=1}^m U(N_i)$. In particular, if all the branes are separated from each other, then the gauge group is entirely Abelian, $U(1)^{N}$. The W-bosons and off-diagonal elements $X^I_{i,j}$ then gain a mass which is proportional to the separation distance between the branes,
\begin{align}\label{eqn:masswbosons}
    M_{ij} \sim \frac{1}{l_s^2}|\vec{x}_i-\vec{x}_j|
\end{align}
where $\vec{x}_i$ is the position of the $i$th brane in the $9-p$ transverse dimensions.

Now, in general, this gauge theory is coupled both to an infinite tower of massive open string states, as well as bulk gravity modes which correspond to closed string states. We are interested in isolating just the light open string states from the rest of the sectors of the string theory. We can achieve this by taking a decoupling limit, $l_s\to 0$ while keeping $g_{\mathrm{YM}}$ fixed and finite\footnote{Instead of taking $l_s\to 0$, one could simply work at energies much lower than $1/l_s$.}. When $p<3$, this means that we must also take $g_s\to 0$ as fast as $l_s^{3-p}$ goes to zero; when $p=3$, we can leave $g_s$ fixed and take $l_s\to 0$ independently; and when $p>3$, there are additional subtleties which we will not need to confront in this paper. 

Once this decoupling is performed, one can approach the physics of the world-volume gauge theory using properties of the full string theory. As an example, we note that Type IIB string theory has a conjectural $\textsl{SL}(2,\mathbb{Z})$ duality group, whose element $S=\left(\begin{smallmatrix} 0 & -1 \\ 1 & 0 \end{smallmatrix}\right)$ acts as
\begin{align}
    S: \ l_s^2 \to l_s^2 g_s, \ \ \ g_s\to \frac{1}{g_s}
\end{align}
while simultaneously exchanging D$p$-branes with other kinds of branes. Therefore, assuming the decoupling limit plays well with the duality under consideration, one can examine the world-volume theories of the brane configurations on both sides of the duality and assert that they are dual as field theories. For instance, the D3-brane maps to itself under S-duality (i.e.\ is ``self-dual"), and so it is easy to see that the $\textsl{SL}(2,\mathbb{Z})$ of Type IIB descends to the well-known field theory duality of $\mathcal{N}=4$ super Yang Mills in (3+1)D with the same name. We will use this kind of logic in \S\ref{subsec:planar} to derive dualities of foliated field theories.

In this paper, we will be dealing with webs of \emph{intersecting} branes. In such situations, for every pair (D$p$,D$q$) of D-branes in spacetime, there are degrees of freedom describing open strings stretched between them; we refer to these as D$p$-D$q$ strings. When the pair has a non-empty intersection, there are massless modes which can be thought of as being localized on the intersection. The precise field content supported on the intersection, as well as the decoupling limits needed to isolate a world-volume field theory, will vary from one example to the next, however, a fairly broad class of intersections can be treated as follows. We keep the discussion general by considering $N$ parallel D$p$-branes intersecting $M$ parallel D$q$-branes on an $(n+1)$-dimensional world-volume, where we take without loss of generality
$$ q \geq p \geq n.$$
The D$p$-D$p$ strings produce a $(p+1)$-dimensional SYM theory on the world-volume of the D$p$-branes, as described in previous paragraphs, and the same goes for the D$q$-D$q$ strings. If we want to retain the gauge theory on the D$p$-branes, the limit in which it decouples from the bulk string theory is $l_s \rightarrow 0$
taken in such a way that $g^2_{\mathrm{YM}}$ remains finite. Restricting to the case that $p\leq 3$, the important point that follows from Eq.\ \eqref{eqn:gym} is that if we take the limit keeping the D$p$ gauge coupling fixed, the couplings of the $(q+1)$-dimensional fields on the D$q$-branes automatically go to zero as long as $q > p$, and so these fields become free and decouple. The world-volume gauge symmetry on the D$q$-branes then becomes a global symmetry from the point of view of the interacting theory on the D$p$-branes. 

There will also be fields living on the $(n+1)$-dimensional intersection which capture the dynamics of the D$p$-D$q$ strings. As we have already said, the precise field content depends on the nature of the intersection, e.g.\ the amount of supersymmetry it preserves. In this work, we will mostly focus on intersections which are ``4ND'' which by definition means  that there are in total 4 directions that are spanned by one of the branes but not the other. In such cases, the joint D$p$/D$q$ configuration preserves 8 real supercharges, and the intersection supports a defect hypermultiplet\footnote{Hypermultiplets are supersymmetric multiplets which arise in theories with 8 supercharges. They contain only complex scalars and fermions.}. This hypermultiplet is in a bifundamental representation with respect to the $U(N)$ \emph{gauge} symmetry of the D$p$-branes, and the $U(M)$ \emph{global} symmetry of the D$q$-branes.

As an example, the full Lagrangian is best understood in the supersymmetric D3/D5 system \cite{Karch:2000gx,DeWolfe:2001pq} corresponding to $q=5$, $p=3$, and $n=2$. In this case the full action can be taken to be schematically of the form
\begin{align}
    S \sim \frac{1}{g^2_{\mathrm{YM}}}(S_B + S_F)
\end{align} 
where $g^2_{\mathrm{YM}}$ is the gauge coupling of the fields living on the stack of D3-branes, $S_B$ is the action of these (3+1)-dimensional fields, and $S_F$ is the action of the (2+1)-dimensional fields living on the defect. By the arguments above, the (5+1)D fields on the D5-brane world-volume decouple, but as long as we keep $g^2_{\mathrm{YM}}$ finite, the (2+1)D and (3+1)D fields remain interacting. The system describes a non-trivial defect conformal field theory.

\section{Foliated field theories }\label{sec:fft}

In this section, 
we will construct D-brane models which naturally hand us foliated field theories. We build on a corpus of existing work, offering a few analogies and re-interpretations along the way that we hope will help bridge the parts of the high-energy and condensed-matter communities that are interested in fractons and exotic symmetry.

We will orient ourselves in \S\ref{subsec:warmupfoliatedgaugetheory} with the simplest example of a foliated field theory and its corresponding D-brane realization. In \S\ref{subsec:xcube}, we will review the X-cube model and explain an analogy to the models we consider in \S\ref{subsec:linearfoliated}-\S\ref{subsec:planar}. After this, we will move on in \S\ref{subsec:linearfoliated} to studying $\textsl{U}(1)$ linear subsystem symmetries in (2+1)D, both in the context of a non-supersymmetric prototype model (in \S\ref{subsubsec:linearfoliatednonsusy}), and also in a brane model  which realizes a supersymmetric version of this prototype (in \S\ref{subsec:D2D4D4}). Having understood the basic idea of the construction, we will repeat it in \S\ref{subsec:planar} for $\textsl{U}(1)$ planar subsystem symmetries in (3+1)D, this time introducing dualities into the mix. 

\subsection{Warm up: foliated gauge theory from stacks of D-branes}\label{subsec:warmupfoliatedgaugetheory}

In general, the foliated field theories we will work with feature two kinds of fields, which we will refer to as ``foliation fields'' and ``bulk fields''. Heuristically, foliation fields behave like stacks of lower-dimensional fields living on the leaves of a foliation of spacetime, whereas bulk fields are ordinary fields which live throughout the bulk of spacetime. Before getting to more sophisticated models in which both kinds of fields are coupled together, it will be useful to warm up with the example of a foliated $\textsl{U}(1)$ gauge theory which just contains the former kind (cf.\ \S 2.2 of \cite{hsin2021comments}). This will give us the opportunity to explain the basics of what a foliated field theory is in the simplest setting. Moreover, from the perspective of string theory, it corresponds to the most naive setup one might imagine implementing: namely, an evenly spaced stack of D-branes.

First, let us describe how to specify a foliation, following \cite{slagle2019foliated,slaglefoliated1,hsin2021comments}. In this paper, all of our examples will feature foliations of space by codimension-1 leaves. In this case, we can specify a foliation with a one--form background foliation field $e(x) = e_\mu(x)dx^\mu$, which by definition points orthogonally to the leaves of the foliation. That is, the tangent vectors $v$ to the leaves satisfy $e(v)=0$. Consistency then requires that $e$ is never zero, and that $e\wedge de = 0$. For simplicity, we will always assume that our foliation fields are closed, $de=0$, so that they trivially satisfy this second constraint. 
If there are $N_{\mathrm{F}}$ different foliations, we will have $N_{\mathrm{F}}$ different foliation fields $e^k(x)$ for $k=1,\dots,N_{\mathrm{F}}$. 

Throughout this section, we will write down non-supersymmetric prototype models, and (less explicitly) supersymmetric variants of them coming from D-branes. We will attempt to be general when it is possible, while we are treating the non-supersymmetric prototypes. However, for our D-brane models, we will essentially always take the bulk spacetime on which our foliated field theory lives to be $\mathbb{R}\times (S^1_R)^d$, where $\mathbb{R}$ represents time and $S^1_R$ is a circle of radius $R$, so that space is a flat $d$-dimensional torus; furthermore, our background foliation fields will be taken to be trivial, i.e.\  $e^k(x) = dx^k$ for $k=1,\dots, N_{\mathrm{F}}$ with $N_{\mathrm{F}}\leq d$. This specialization is in order to guarantee supersymmetry of our brane constructions. 

\begin{table}
\begin{center}
\begin{tabular}{c|cccc|cccccc}
&$t$&$x$&$y$&$z$&$4$&$5$&$6$&$7$&$8$&$9$ \cr
\hline
$\mathrm{D}2^I$ & \textbf{x}&\textbf{x}&\textbf{x}&::&&&&&& \\
\end{tabular}
\caption{A diagram encoding the layout of the D2$^I$-branes in spacetime. An \textbf{x} denotes a direction along which the branes extend, and :: a direction along which they are localized but form an evenly spaced lattice with spacing $\delta$. If an entry is empty, the brane is at the origin of the corresponding dimension, i.e.\ at $x^i=0$.}\label{fig:D2stack}
\end{center}
\end{table}

With these preliminaries in place, let us define our first example of a foliated field theory. Let $e^k$ for $k=1,\dots, N_{\mathrm{F}}$ be a collection of background foliation fields in $(d+1)$ spacetime dimensions. Define a foliated $\textsl{U}(1)$ gauge field $B^k$ to be a two-form which satisfies $B^k\wedge e^k=0$ and which enjoys a gauge transformation of the form $B^k\to B^k+d\lambda^k$, where $\lambda^k$ is a one-form satisfying $\lambda^k\wedge e^k = 0$. We take the action of the $B^k$ to be
\begin{align}\label{eqn:foliatedgaugetheory}
    \mathcal{L} =- \frac{1}{2g_f^2}\sum_{k=1}^{N_{\mathrm{F}}} dB^k \wedge \star dB^k.
\end{align}
where $g_f$ is a coupling constant. To gain intuition for these definitions, let us specialize our spacetime to be $\mathbb{R}^4$ and consider the case of a single trivial foliation field $e = dz$. In that case, the constraint $B\wedge dz=0$ tells us that we can write $B=\widetilde{B}\wedge dz$ with $\widetilde{B} =\widetilde{B}_t dt+  \widetilde{B}_x dx +\widetilde{B}_y dy$ a one-form which has no $z$-component. Similarly, the constraint $\lambda\wedge dz=0$ implies that the gauge transformation parameter can be written as $\lambda = \widetilde{\lambda}dz$ with $\widetilde{\lambda}$ a scalar function, and the gauge transformation simply becomes $\widetilde{B}_i \to \widetilde{B}_i+\partial_i\widetilde{\lambda}$. Plugging $B=\widetilde{B}\wedge dz$  into Eq.\ \eqref{eqn:foliatedgaugetheory} then reveals that
\begin{align}\label{eqn:foliatedgaugetheorytrivialbackground} 
    \mathcal{L} = -\frac{1}{4g_f^2}\sum_{i,j=t,x,y} \widetilde{F}_{ij}\widetilde{F}^{ij}, \ \ \ \ \ \ \widetilde{F}_{ij} = \partial_i\widetilde{B}_j - \partial_j \widetilde{B}_i.
\end{align}
In other words, $\mathcal{L}$ is simply describing a decoupled stack of (2+1)D $\textsl{U}(1)$ gauge fields in the $z$-direction, each with an ordinary free Maxwell action. 

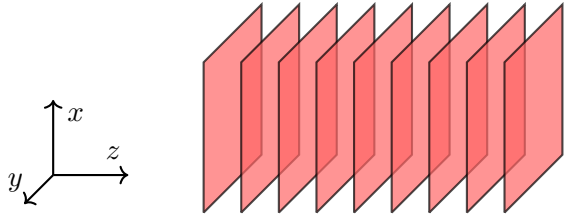
\begin{figure}
    \centering
\begin{tikzpicture}
\foreach \x in {0,...,8}
{
\fill[red!60,draw=black,thick, opacity=.7] (\x*.5,0,0) -- (\x*.5,0,-2) -- (\x*.5,2,-2) -- (\x*.5,2,0) --(\x*.5,0,0);
}
\draw[thick,->] (-2,.5,0) -- (-2,1.5,0);
\node[] at (-1.2,.8) {$z$};
\draw[thick,->] (-2,0.5,0) -- (-2,0.5,1);
\node[] at (-1.7,1.3) {$x$};
\draw[thick,->] (-2,0.5,0) -- (-1,0.5,0);
\node[] at (-2.5,.4) {$y$};
\end{tikzpicture}
    \caption{A stack of D2-branes with spacing $\delta$ in Type IIA string theory.}
    \label{fig:D2branestack}
\end{figure}

In the case of a single background foliation field $e=dz$, we can achieve a supersymmetric version of this foliated gauge theory from D-branes as follows. Simply consider Type IIA string theory with the $x,y,z$ directions compactified on circles of radius $R$. Take a stack of $L$ evenly spaced D2-branes which span the $t,x,y$ directions and form a lattice with spacing $\delta$ in the $z$-direction, so that $2\pi R = L\delta$.  We label the $I$th brane D2$^I$. As described in \S\ref{sec:branereminders}, the theory on their world-volumes is a (2+1)D $\textsl{U}(L)$ $\mathcal{N}=8$ super Yang-Mills theory, with its gauge group spontaneously broken down to $\textsl{U}(1)^L$. We call $\widetilde{A}^I$ the $\textsl{U}(1)$ gauge field on the $I$th D2-brane, and $\widetilde{F}^I$ its field strength. See Figure \ref{fig:D2branestack} and Table \ref{fig:D2stack} for a summary of this setup.

We would like to take a large $L$ (and hence $\delta\to 0$) limit and interpret these $\textsl{U}(1)$ gauge fields as a foliated stack of decoupled (2+1)D free Maxwell theories, as in Eq.\ \eqref{eqn:foliatedgaugetheorytrivialbackground}. The one subtlety is that the stacks are in fact coupled together by strings which stretch between the different layers. These are the massive W-bosons associated with the spontaneous breaking of the gauge group.  To make sure that they do not contribute, we must take $\delta, l_s\to 0$ simultaneously in such a way that the mass of the W-bosons, which goes like $M_{\mathrm{W}} \sim \delta/l_s^2$ (cf.\  Eq.\ \eqref{eqn:masswbosons}), stays very large. If we were only interested in cooking up a theory of a large $L$ number of decoupled (2+1)D $\textsl{U}(1)$ gauge fields (plus their super partners),
\begin{align}\label{eqn:decoupleddiscretestack}
S \sim -\frac{1}{4g_{\mathrm{YM}}^2}\sum_{I=1}^L\int d^3x \widetilde{F}^{I}_{ij}\widetilde{F}^{Iij} + (\text{super partners}),
\end{align}
then we would simply take $g_s\to 0$, in conjunction with $\delta,l_s\to 0$ as described above, so that $g_{\mathrm{YM}}^2 \sim g_s/l_s$ remains fixed and finite.

However, we may also try to make closer contact with the FFT description. This would involve taking the continuum limit in such a way that the sum over $I$ in Eq.\ \eqref{eqn:decoupleddiscretestack} could be replaced with an integral over an emergent dimension, 
\begin{align}
    \delta \sum_{I=1}^L \xrightarrow{\substack{L\to \infty \\ \delta \to 0}} \int dz.
\end{align}
In order to achieve this, we multiply Eq.\ \eqref{eqn:decoupleddiscretestack} by $1 = \delta/\delta$. The $\delta$ in the numerator is used in converting the sum to an integral, whereas the $\delta$ in the denominator appears as a proportionality constant relating the coupling of the foliated field theory to the coupling on the world-volume of the D2-branes,
\begin{align}
    g_f^2=g_{\mathrm{YM}}^2\delta \sim \delta g_s/l_s.
\end{align} 
Thus,  it is clear that we should arrange for a slightly different decoupling limit than the one described in the previous paragraph: we take $g_s,l_s,\delta\to 0$ again, keeping the W-bosons heavy as before, but this time such that the foliated gauge coupling $g_f^2 \sim \delta g_s/l_s$ remains fixed and finite (as opposed to the Yang-Mills coupling). This ensures that the action Eq.\ \eqref{eqn:decoupleddiscretestack} goes over to the Lagrangian in Eq.\ \eqref{eqn:foliatedgaugetheorytrivialbackground} in the continuum limit. In addition, the bosonic content of the theory will also include 7 foliated scalars (cf.\ Eq.\ \eqref{eqn:bosonicSYM}) which are neutral with respect to the $\textsl{U}(1)$ foliated gauge theory, and whose action can be determined similarly. One can straightforwardly dualize the foliated gauge field by lifting to M-theory, where it becomes a foliated scalar describing the embedding of the M2-brane stack into the 11th dimension.

\subsection{Intermezzo: the X-cube model}\label{subsec:xcube}

In the previous subsection, we treated an example of a theory which only had foliation fields, i.e.\ no bulk fields. We saw that its physics was essentially the same as that of a stack of trivially decoupled lower-dimensional theories. One way that one could imagine producing a more interesting system is to  couple the layers of such a theory together by immersing them inside of a bulk  whose fields permeate all of space. This is  what we will do in the next two subsections, \S\ref{subsec:linearfoliated} and \S\ref{subsec:planar}.

But first, in order to motivate this approach, we will illustrate how it is analogous to a particular coupled layer construction of the X-cube Hamiltonian \cite{slagle2019foliated} (see also \cite{ma2017fracton} for a related model). Much of the perspective we take in this subsection was emphasized recently in \cite{qi2021fracton,brdw}. Consider a 3d $L_x\times L_y\times L_z$ cubic lattice $\Lambda$, and place a qubit $\mathbb{C}^2$ on each edge. The Hamiltonian of the X-cube lattice model is 
\begin{align}
    H_{\mathrm{XC}} = - \sum_{\substack{\mathrm{vertices}\\v}}\left(
\begin{tikzpicture}[baseline=-.65ex]
\draw[red,line width=.5mm] (-.75,0,0) -- (.75,0,0);
\draw (0,-.75,0) -- (0,.75,0);
\draw[red,line width=.5 mm] (0,0,-.75) -- (0,0,.75);
\node at (0.14,-.15,0) {$\scriptstyle v$};
\end{tikzpicture}
+
\begin{tikzpicture}[baseline=-.65ex]
\draw (0,0,-.75) -- (0,0,.75);
\draw[red,line width=.5 mm] (-.75,0,0) -- (.75,0,0);
\draw[red,line width=.5mm] (0,-.75,0) -- (0,.75,0);
\node at (0.14,-.15,0) {$\scriptstyle v$};
\end{tikzpicture}
+
\begin{tikzpicture}[baseline=-.65ex]
\draw (-.75,0,0) -- (.75,0,0);
\draw[red,line width=.5mm] (0,-.75,0) -- (0,.75,0);
\draw[red,line width=.5 mm] (0,0,-.75) -- (0,0,.75);
\node at (0.14,-.15,0) {$\scriptstyle v$};
\end{tikzpicture}
\right)
-\sum_{\substack{\mathrm{cubes}\\c}}
\begin{tikzpicture}[scale=1,fill opacity=0.4,line cap=round,line join=round,baseline=3.5ex]
\draw (-.25,0,0) -- (1.25,0,0);
\draw (-.25,1,0) -- (1.25,1,0);
\draw (-.25,0,-1) -- (1.25,0,-1);
\draw (-.25,1,-1) -- (1.25,1,-1);
\draw (0,-.25,0) -- (0,1.25,0);
\draw (1,-.25,0) -- (1,1.25,0);
\draw (1,-.25,-1) -- (1,1.25,-1);
\draw (0,-.25,-1) -- (0,1.25,-1);
\draw (0,0,-1.25) -- (0,0,.25);
\draw (0,1,-1.25) -- (0,1,.25);
\draw (1,0,-1.25) -- (1,0,.25);
\draw (1,1,-1.25) -- (1,1,.25);
\draw[yellow,line width=.5mm] (0,0,0) -- (1,0,0);
\draw[yellow,line width=.5mm] (0,0,-1) -- (1,0,-1);
\draw[yellow, line width=.5mm] (0,0,-1) -- (0,0,0);
\draw[yellow, line width=.5mm] (1,0,-1) -- (1,0,0);
\draw[yellow,line width=.5mm] (0,1,0) -- (1,1,0);
\draw[yellow,line width=.5mm] (0,1,-1) -- (1,1,-1);
\draw[yellow, line width=.5mm] (0,1,-1) -- (0,1,0);
\draw[yellow, line width=.5mm] (1,1,-1) -- (1,1,0);
\draw[yellow, line width=.5mm] (1,0,0) -- (1,1,0);
\draw[yellow, line width=.5mm] (1,0,-1) -- (1,1,-1);
\draw[yellow, line width=.5mm] (0,0,-1) -- (0,1,-1);
\draw[yellow, line width=.5mm] (0,0,0) -- (0,1,0);
\end{tikzpicture}
\end{align}
where the red/yellow lines indicate a Pauli X/Pauli Z operator applied to the corresponding edge. This model has what one might call a ``$\mathbb{Z}_2$ one-form planar subsystem symmetry''. What this means essentially is that within each plane, the model admits operators which behave like the symmetry operators of an ordinary $\mathbb{Z}_2$ one-form symmetry in (2+1)D (see \cite{gaiotto2015generalized} for a lucid discussion of ordinary higher-form symmetries). Indeed, it is straightforward to check that, for each plane $P$ and each closed loop $\tilde{\gamma}$ through the \emph{dual} lattice of the 2d sublattice picked out by $P$, the operator 
\begin{align}
    {^\mathrm{XC}}U^P_{\tilde{\gamma}} := \prod_{e\in\tilde{\gamma}} X_e
\end{align}
commutes with $H_{\mathrm{XC}}$, where the product is over edges which are perpendicularly bisected by the path $\tilde{\gamma}$. Moreover, the operators are topological in the sense that one can deform the path $\tilde{\gamma}$ \emph{within} the plane $P$ to which it belongs without changing how it acts (so long as one is careful not to pass through charged operators). 

One can approximate the symmetry structure described in the previous paragraph by considering three orthogonal decoupled stacks of (2+1)D toric code layers, which we again think of as forming a 3d cubic lattice $\Lambda$, this time with two qubits per edge (since each edge belongs to two toric code planes). Indeed, the (2+1)D toric code famously has an ordinary $\mathbb{Z}_2$ one-form symmetry\footnote{It actually has two, an electric one-form symmetry and a magnetic one-form symmetry, but only the latter is relevant for this discussion.}, and so three decoupled stacks of toric code layers will trivially have a planar $\mathbb{Z}_2$ one-form subsystem symmetry. More precisely, the symmetry operators are 
\begin{align}\label{eqn:TCsymmetryops}
    {^\mathrm{TC}}U^P_{\tilde{\gamma}} = \prod_{e\in\tilde{\gamma}} X_e^P
\end{align}
where now, since each edge of the 3d cubic lattice has two qubits, we have specified which of the two we are acting on by including the plane $P$ in the superscript of the Pauli X operator. 

However, there is one crucial difference between the symmetries of the X-cube model and the symmetries of decoupled stacks of toric code layers, which one can see as follows. Note that, in the case of the toric layers, one can combine the symmetry operations on the various planes to produce an ordinary (3+1)D $\mathbb{Z}_2$ one-form symmetry, which one might call the \emph{diagonal} one-form symmetry subgroup. The symmetry operators of a one-form symmetry in (3+1)D are supported on surfaces; in this present case, they are supported on closed membranes $\tilde{m}$ formed out of plaquettes of the dual lattice $\Lambda^\ast$. Each such membrane $\tilde{m}$ intersects each 2d plane $P$ in some (possibly zero) number of paths $\tilde{\gamma}$. With this in mind, we can then build up a membrane operator $V_{\tilde{m}}$ out of the line operators from Eq.\ \eqref{eqn:TCsymmetryops} as
\begin{align}\label{eqn:diagonaloneform}
    {^\mathrm{TC}}V_{\tilde{m}} \sim \prod_P \prod_{\tilde{\gamma} \in P \cap \tilde{m}} {^\mathrm{TC}}U^P_{\tilde{\gamma}}.
\end{align}
These operators commute with the Hamiltonian and are topological simply because the operators ${^\mathrm{TC}}U_{\tilde{\gamma}}^P$ have the same properties. One can attempt to form a similar membrane operator in the X-cube model, however one discovers that it is trivial,
\begin{align}
    {^\mathrm{XC}}V_{\tilde{m}} \sim \prod_P \prod_{\tilde{\gamma}\in P\cap \tilde{m}} {^\mathrm{XC}}U_{\tilde{\gamma}}^P = 1.
\end{align}
The reason for this is that every edge $e$ which is intersected by the membrane $\tilde{m}$ belongs to two planes, $P_1$ and $P_2$. In the case of the toric layers, the symmetry operator ${^\mathrm{TC}}V_{\tilde{m}}$ thus receives a contribution $X_e^{P_1} X_e^{P_2}$ from each such edge. However in the case of the X-cube model, there is only one qubit per edge, and so the analogous contribution is $X_e X_e = 1$. 

Intuitively, the difference between the X-cube model and the decoupled toric code layers is that the former is subjected to some kind of global relation between the one-form symmetries on the different layers which renders the diagonal one-form symmetry group trivial. One can impose this relation on the toric code layers simply by \emph{gauging} the diagonal one-form symmetry group of Eq.\ \eqref{eqn:diagonaloneform}, or more suggestively, by immersing the (2+1)D toric code layers in a (3+1)D $\mathbb{Z}_2$ two-form gauge theory (which on the lattice is described by the (3+1)D toric code). This procedure is indeed effective in converting the toric code layers into the X-cube model. In the continuum, it leads to a foliated field theory description of the X-cube model \cite{slagle2019foliated},
\begin{align}
    \mathcal{L} \sim \frac{1}{\pi}\left[b\wedge da+ \sum_{k=1}^3  (e^k\wedge B^k \wedge dA^k - e^k\wedge b\wedge A^k)  \right].
\end{align}
In the above, $b$ is a two-form field; $a$, $B^k$, and $A^k$ are all one-form fields; and for simplicity, we can take the foliation fields to be trivial, $e^k = dx^k$. The first term is a continuum BF description of $\mathbb{Z}_2$ two-form  gauge theory in (3+1)D (see e.g.\ \cite{kapustin2014coupling}), the second term describes three orthogonal stacks of toric code layers (using the Chern-Simons description of the toric code), and the last term couples the the toric code stacks to the bulk gauge theory. 

The idea illustrated above is general. For example, one can show that the (2+1)D plaquette Ising model (see \cite{johnston2017plaquette} for a review), a theory with $\mathbb{Z}_2$ linear zero-form subsystem symmetry, can be obtained by taking stacks of decoupled Ising model wires and immersing them in a conventional $\mathbb{Z}_2$ gauge theory \cite{brdw} to impose a global relation between the symmetries on the different wires. In the next subsections, we will apply a similar approach to $\textsl{U}(1)$ zero-form subsystem symmetries.

\subsection{Linear $\normalfont \textsl{U}(1)$ subsystem symmetry in (2+1)D}\label{subsec:linearfoliated}
With the discussion of the previous subsection in mind, in the rest of \S\ref{sec:fft}, we will study slightly more complicated brane setups which realize FFTs which have both bulk gauge fields and also foliated matter fields. Let us briefly summarize how the construction will go. In this paragraph we will take all groups to be $\textsl{U}(1)$, though we will be slightly more general in the subsequent sections. If we are aiming for a $(p+1)$--dimensional theory, we first place a ``bulk brane'' $^{\mathrm{B}}$D$p$ in spacetime. The strings which have both endpoints anchored on this bulk brane are captured by a $(p+1)$-dimensional supersymmetric $\textsl{U}(1)$ gauge theory on its world-volume, as we have described in \S\ref{sec:branereminders}. We then introduce $N_{\mathrm{F}}$ families of ``foliation branes'' ${^\mathrm{F}}$D$q^{(k)}_\alpha$, where $k=1,\dots,N_{\mathrm{F}}$, and $\alpha=1,\dots,L_k$ runs over the $L_k$ leaves of the $k$th foliation. For the most part in this section, we take $q=p+2$. We orient each foliation brane in spacetime so that it has a 4ND and codimension-1 intersection with the bulk brane; strings which stretch from a given foliation brane ${^\mathrm{F}}$D$q^{(k)}_\alpha$ to the bulk branes $^{\mathrm{B}}$D$p$ therefore contribute a $p$-dimensional defect theory which is localized on their intersection, and which, with the $^{\mathrm{B}}$D$p$ gauge coupling fixed, carries a $\textsl{U}(1)$ global symmetry (see \S\ref{sec:branereminders}). Following the discussion at the end of \S\ref{sec:branereminders}, the rest of the string sectors decouple, leaving in total a supersymmetric foliated field theory which has a $\left(\prod_{k=1}^{N_\mathrm{F}}\textsl{U}(1)^{L_k}\right)\big/\textsl{U}(1)$ subsystem symmetry group. Specializing to $p=3$ and $q=5$ in this construction, which we will do in \S\ref{subsec:planar}, puts us in the realm of brane webs of Hanany--Witten type. Building on the results of \cite{hanany1997type,gaiotto2009supersymmetric,gaiotto2009s,seiberg2016duality,Karch:2016sxi}, we find that the Hanany--Witten program --- that is, using S--duality of Type IIB string theory to explain mirror symmetry of (2+1)D $\mathcal{N}=4$ gauge theories --- extends naturally to certain (3+1)D foliated field theories with planar subsystem symmetries.

\subsubsection{Non-supersymmetric prototypes}\label{subsubsec:linearfoliatednonsusy}

We start by writing down toy versions of the (2+1)D foliated field theory which we will eventually realize on branes in \S\ref{subsec:D2D4D4}. Unlike the brane theory, the toy models in this section are simpler in that they are non-supersymmetric, and they have the minimal field content required to exhibit the main idea of the construction.

Fix $N_{\mathrm{F}}$ background foliation fields $e^k$. Once the foliations are chosen, we would like to place a theory with a $\textsl{U}(1)$ global symmetry on each leaf. In this simple example, our leaves are wires which foliate a spatial $T^2$, and so it is natural to place a (1+1)D free compact boson on each wire. We can achieve this by introducing periodic scalars $\varphi^k(t,x,y)$, with $k=1,\dots,N_{\mathrm{F}}$ running over the different foliations,  and taking their action to be 
\begin{align}\label{FFTcompactboson}
\begin{split}
    \mathcal{L} &= \frac{1}{4\pi}\sum_{k=1}^{N_{\mathrm{F}}}\frac{R_k^2}{2} d\widehat{\varphi}^k\wedge \star d\widehat{\varphi}^k,   \ \ \ \ \ \ \ (\widehat{\varphi}^k \equiv \varphi^k \wedge e^k) \\
\end{split}
\end{align}
where $\star$ denotes the hodge star operator. Intuition for this Lagrangian can once again be gained by considering the case of a trivial foliation field, e.g.\ $e = dy$ (corresponding to $e_\mu =\delta^y_\mu$), in which case 
the action essentially becomes that of a (1+1)D compact boson in the $t$,$x$ directions, but integrated over the $y$ direction as well. This motivates us to think of $R_k$ as roughly the target space radius of the free bosons living on the leaves of the $k$th foliation. If one would like to work in some level of generality, it is natural to allow $R_k$ to vary in the direction orthogonal to the leaves of the $k$th foliation, while requiring it to be constant along directions parallel to the leaves. We have repackaged $\varphi^k$ (a 0--form field) into a 1-form field $\widehat{\varphi}^k$ for ease of notation. If one would like, one could dispense with $\varphi^k$ and work entirely with $\widehat{\varphi}^k$ from the outset by supplementing the Lagrangian in Eq.\ \eqref{FFTcompactboson} with a constraint $\widehat{\varphi}^k\wedge e^k = 0$, similar to what was done in the previous section for the foliated gauge theory of $B^k$. Finally, we note that this Lagrangian has a shift symmetry which acts as 
\begin{align}
   \varphi^k\to \varphi^k+c^k, \ \ \ 
\end{align}
where $c^k$ can depend on spacetime, so long as $dc^k\wedge e^k=0$. In other words, we demand that $c^k$ is constant along the leaves of the $k$th foliation, while allowing it to vary from leaf to leaf; this is an example of a subsystem symmetry. We note the similarity of this model to the foliated gauge theory presented in \S 2.2 of \cite{hsin2021comments}.

So far, this theory is trivial in the sense that it simply consists of a continuum of decoupled lower-dimensional theories. We can obtain a more interesting theory by coupling it to conventional (2+1)D electrodynamics, in analogy with the X-cube model described in the previous subsection, \S\ref{subsec:xcube}. We will do this by gauging the overall $\textsl{U}(1)$ which rotates the scalars on all the wires by the same amount. Doing this, we find 
\begin{align}\label{eqn:foliatednonsusy2p1d} 
    \mathcal{L} = -\frac{1}{2e^2}F\wedge \star F + \frac{1}{4\pi}\sum_{k=1}^{N_{\mathrm{F}}}\frac{R_k^2}{2}(d\widehat{\varphi}^k-\widehat{A}^k)\wedge \star(d\widehat\varphi^k-\widehat{A}^k), \ \ \ \ \widehat{A}^k \equiv A\wedge e^k,
\end{align}
where $A$ is a (2+1)D $\textsl{U}(1)$ gauge field, and $F=dA$ is its field strength. In terms of our previously advertised terminology, we will refer to fields like $\varphi^k$ as foliation fields, and fields like $A$ as bulk fields.

We point out in passing that this model has two kinds of dualities. The first is obtained essentially by T--dualizing the bosons on the leaves of the foliations. The second involves dualizing the bulk photon to a compact scalar. We postpone a more detailed discussion of dualities to the setting of (3+1)D planar subsystem symmetries in \S\ref{subsec:planar}, where we will see that the composition of these two kinds of dualities can be understood as a consequence of S-duality of Type IIB string theory.

Our brane model in the next section will feature a subsystem symmetry that is realized linearly, as opposed to the action in Eq.\ \eqref{eqn:foliatednonsusy2p1d}, where it is realized as a shift symmetry. We can achieve this in the present setting by promoting the compact scalars $\varphi^k$ to complex scalars $\Phi^k$. Then for example, we can take the action to be 
\begin{align}\label{eqn:complexfftnonsusy2p1d}
    \mathcal{L} = -\frac{1}{2e^2}F\wedge \star F + \sum_{k=1}^{N_{\mathrm{F}}} (\mathcal{D}\widehat\Phi^k)^\ast \wedge \star \mathcal{D}\widehat\Phi^k-V(|\Phi^k|^2)e^k\wedge\star e^k, \ \ \ \ \ \widehat\Phi^k\equiv \Phi^k\wedge e^k
\end{align}
where $\mathcal{D} = d-iA$ is a covariant derivative. This theory then has a $\textsl{U}(1)$ subsystem symmetry of the form 
\begin{align}
    \Phi^k \to e^{ic^k}\Phi^k
\end{align}
where again, $dc^k\wedge e^k=0$. If one chooses $V(|\Phi^k|^2) \sim -\mu |\Phi^k|^2+\lambda |\Phi^k|^4$ and expands $\Phi^k = \rho^k e^{i\varphi^k}$, then the action Eq.\ \eqref{eqn:complexfftnonsusy2p1d} flows to Eq.\ \eqref{eqn:foliatednonsusy2p1d} in the IR.

\subsubsection{A supersymmetric Type IIA brane model}\label{subsec:D2D4D4}

\begin{table}
\begin{center}
\begin{tabular}{c|ccc|ccccccc}
&$t$&$x$&$y$&$3$&$4$&$5$&$6$&$7$&$8$&$9$ \cr
\hline
${^\mathrm{B}}\mathrm{D}2$ & \textbf{x}&\textbf{x}&\textbf{x}&&&&&&& \\
${^\mathrm{F}}\mathrm{D}4^{(y)}_\alpha$ & \textbf{x}&\textbf{x}& :: &\textbf{x}&\textbf{x}&\textbf{x}&&&& \\
${^\mathrm{F}}\mathrm{D}4^{(x)}_\alpha$ & \textbf{x}& :: &\textbf{x}&\textbf{x}&\textbf{x}&&\textbf{x}&&&
\end{tabular}
\caption{A diagram encoding the layout of our branes in spacetime. An \textbf{x} denotes a direction along which the branes extend, and :: a direction along which they are localized but form an evenly spaced lattice with spacing $\delta$. If an entry is empty, the brane is at the origin of the corresponding dimension, i.e.\ at $x^i=0$.}\label{fig:D2D4D4}
\end{center}
\end{table}

\begin{figure}
    \centering
\begin{tikzpicture}
\fill[blue!30] (-.7,-.7) rectangle (1.6*3+.7,1.6*3+.7);
\foreach \x in {0,...,3}
    {\draw (-.7,1.6*\x) -- (1.6*3+.7,1.6*\x);
    \draw (-.7,1.6*\x+.08) -- (1.6*3+.7,1.6*\x+.08);
    \draw (-.7,1.6*\x-.08) -- (1.6*3+.7,1.6*\x-.08);
    \draw (1.6*\x,-.7) -- (1.6*\x,1.6*3+.7);
    \draw (1.6*\x+.08,-.7) -- (1.6*\x+.08,1.6*3+.7);
    \draw (1.6*\x-.08,-.7) -- (1.6*\x-.08,1.6*3+.7);
    }
\node[] at (1.6*3+1.3,0) {${^\mathrm{F}}$D4$_{L}^{(y)}$};
\node[] at (1.6*3+1.3,1.6) {${^\mathrm{F}}$D4$_{1}^{(y)}$};
\node[] at (1.6*3+1.3,1.6*2) {${^\mathrm{F}}$D4$_{2}^{(y)}$};
\node[] at (1.6*3+1.3,1.6*3) {${^\mathrm{F}}$D4$_{3}^{(y)}$};
\node[] at (1.6*3+1.3,1.6*3+1) {$\vdots$};
\node[] at (0,-1) {${^\mathrm{F}}$D4$_{L}^{(x)}$};
\node[] at (1.6*1,-1) {${^\mathrm{F}}$D4$_{1}^{(x)}$};
\node[] at (1.6*2,-1) {${^\mathrm{F}}$D4$_{2}^{(x)}$};
\node[] at (1.6*3,-1) {${^\mathrm{F}}$D4$_{3}^{(x)}$};
\node[] at (1.6*3+1.2,-1) {$\cdots$};
\node[] at (1.6+1.6/2,1.6+1.6/2) {${^\mathrm{B}}$D2};
\draw[->] (-2.2,.4) -- (-2.2+.9,.4) node[anchor= north] {$x$};
\draw[->] (-2.2,.4) -- (-2.2,1.3) node[anchor= east] {$y$};
\node[] at (-1.1,1.6*2) {$M \ \{$};
\end{tikzpicture}
    \caption{A cross-section of the branes in the $x$-$y$ plane. The blue shaded region denotes the $N$ ${^\mathrm{B}}$D2-branes which wrap the entire plane. The grid is formed by the foliation branes. Each row/column is a spindle of $M$ branes; we have drawn these branes with a slight separation for visual clarity, but in the actual brane setup they are directly on top of each other.}
    \label{fig:branelayoutD2D4D4}
\end{figure}
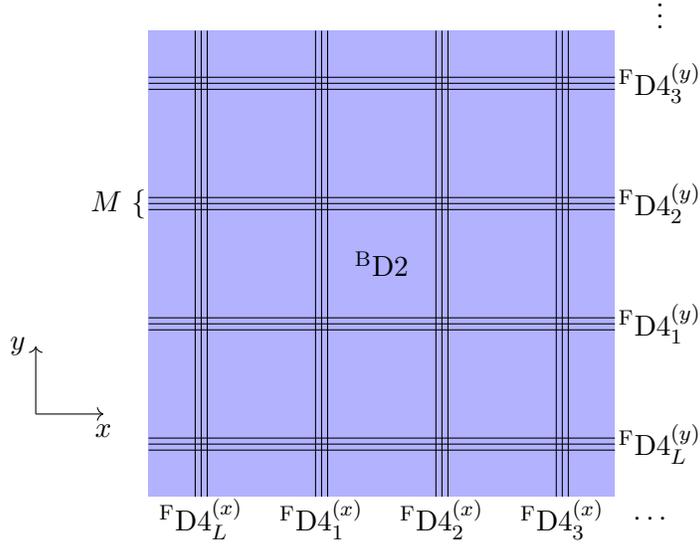

In this section, we will argue that a supersymmetric version of the foliated field theory in Eq.\ \eqref{eqn:foliatednonsusy2p1d} (with trivial background foliation fields $e^1=dx^1$ and $e^2=dx^2$) can be realized on the world-volumes of intersecting D-branes. We work in Type IIA string theory on $\mathbb{R}^{1,7}\times S^1_{x}\times S^1_{y}$, where the two circle factors in spacetime both have radius $R$ and are coordinatized by $x\cong x+2\pi R$ and $y\cong y+2\pi R$. Our brane construction proceeds by first wrapping $N$ ${^\mathrm{B}}$D2-branes along these two circle factors; we take all of them to be located at the origin of the remaining seven spatial dimensions. The ``B'' in the superscript stands for ``bulk'', and is meant to emphasize the eventual interpretation of these branes (or rather, the strings whose endpoints are both anchored on these branes) as contributing bulk fields of the foliated field theory. Next, we draw an $L\times L$ lattice with spacing $\delta$ on the $S^1_x\times S^1_y$ two--torus. We think of this as two discrete foliations of $S^1_x\times S^1_y$ by parallel wires. On each wire stretched in the $x$-direction, we place $M$ ${^\mathrm{F}}$D4$_\alpha^{(y)}$-branes in spacetime which are stretched along the $x^3,x^4,x^5$ directions, and intersect the $S^1_x\times S^1_y$ two--torus in the $\alpha$th row.\footnote{In fact, one can place a different number $M_\alpha^{(y)}$ of branes on each wire, but to reduce notational clutter, we demand translational invariance and simply put $M_\alpha^{(y)}=M$ for all $\alpha$.} We likewise define ${^\mathrm{F}}$D4$_\alpha^{(x)}$-branes, which are stretched along the $x^3$, $x^4$, and $x^6$ directions, and intersect $S_x^1\times S_y^1$ in the $\alpha$th column. The ``F'' in the superscript here stands for ``foliation'', and is meant to signify that the strings which stretch from one of the bulk branes to one of the foliation branes will ultimately give rise to lower-dimensional foliation fields of the foliated field theory. This discussion is summarized in Table \ref{fig:D2D4D4} and Figure \ref{fig:branelayoutD2D4D4}. Nothing essential is lost if one would like to specialize in their heads to the case that $N=M=1$.

We view the physics from the perspective of the (2+1)-dimensions $t$, $x$, and $y$. Overall the system has the symmetry of a square lattice. The ${^\mathrm{B}}$D2-${^\mathrm{B}}$D2 strings give rise to a $(2+1)$D $\textsl{U}(N)$ $\mathcal{N}=8$ super Yang-Mills theory, whose bosonic content includes a $\textsl{U}(N)$ gauge field $A$ and $7$ scalars $X^I$ in its adjoint representation, $I=1,\dots, 7$. These are the bulk fields of the foliated field theory.
Every single brane intersection is 4ND; this is true for each of the D4-branes relative to the D2-branes, and also for the two sets of D4-branes relative to each other. This means that the system is supersymmetric, and every brane intersection localizes a hypermultiplet, just like the D3/D5 example from \S\ref{sec:branereminders}. These hypermultiplets are the foliation fields of the foliated field theory. The bosonic content of the defect hypermultiplet coming from e.g.\ ${^\mathrm{B}}$D2-${^\mathrm{F}}$D4$_\alpha^{(y)}$ strings is simply two (1+1)D complex scalars $\Phi_{\alpha}^{(y)},\widetilde{\Phi}_{\alpha}^{(y)}$, one of which ($\Phi_{\alpha}^{(y)}$) transforms in the fundamental of the $\textsl{U}(N)$ on the ${^\mathrm{B}}$D2-brane, while transforming in the anti-fundamental of the $\textsl{U}(M)_\alpha^{(y)}$. The other ($\widetilde{\Phi}_{\alpha}^{(y)}$) is in the anti-fundamental of $\textsl{U}(N)$ while being in the fundamental of $\textsl{U}(M)_\alpha^{(y)}$.  Similar comments apply to the hypermultiplets corresponding to the ${^\mathrm{B}}$D2-${^\mathrm{F}}$D4$_\alpha^{(x)}$ strings.  In the continuum limit, we think of the $\alpha$ index as one of the spatial coordinates on which the  field depends, e.g.\ we interpret

\begin{align}
    \Phi^{(y)}_{\alpha}(t,x),\widetilde{\Phi}^{(y)}_{\alpha}(t,x)\xrightarrow{\mathrm{cont.}} \Phi^{(y)}(t,x,y), \widetilde{\Phi}^{(y)}(t,x,y)
\end{align} 
where $y\sim \alpha\delta$. Hence, we can interpret $\Phi^{(y)},\widetilde{\Phi}^{(y)}$ as foliation fields, corresponding to the trivial foliation $e^{(y)}=dy$. The bosonic field content is summarized in Table \ref{tab:fieldsD2D4D4}.

\begin{table}[h]
    \centering
    \begin{tabular}{c|c c c c}
         & Sector & $\textsl{U}(N)$ & $\textsl{U}(M)_\beta^{(y)}$ & $\textsl{U}(M)_\beta^{(x)}$ \\
         \hline
        $A_\mu$ & ${^\mathrm{B}}$D2-${^\mathrm{B}}$D2 & $N^2$ & & \\
        $X^I$ & ${^\mathrm{B}}$D2-${^\mathrm{B}}$D2 & $N^2$ & & \\
        $\Phi_{\alpha}^{(y)}$ & ${^\mathrm{B}}$D2-${^\mathrm{F}}$D4$^{(y)}_\alpha$ & $N$ & $\delta_{\alpha\beta}\overline{M}$ & \\
        $\widetilde{\Phi}_{\alpha}^{(y)}$ & ${^\mathrm{B}}$D2-${^\mathrm{F}}$D4$^{(y)}_\alpha$ & $\overline{N}$ & $\delta_{\alpha\beta}M$ &  \\
        $\Phi_{\alpha}^{(x)}$ & ${^\mathrm{B}}$D2-${^\mathrm{F}}$D4$_\alpha^{(x)}$ & $N$ & & $\delta_{\alpha\beta}\overline{M}$ \\
        $\widetilde{\Phi}_{\alpha}^{(x)}$ & ${^\mathrm{B}}$D2-${^\mathrm{F}}$D4$_\alpha^{(x)}$ & $\overline{N}$ & & $\delta_{\alpha\beta}M$ \\
    \end{tabular}
    \caption{A summary of the bosonic field content of the brane setup.}
    \label{tab:fieldsD2D4D4}
\end{table}

So far, the fields we have described are precisely the ones we want to keep to produce a foliated field theory. The action which governs them schematically takes the form 
\begin{align}\label{eqn:schematicd2d4d4p}
    S= \frac{1}{g_{\mathrm{YM}}^2} \left(S_B + \sum_{\alpha=1}^L (S_{F,\alpha}^{(x)} + S_{F,\alpha}^{(y)})  \right)
\end{align}
where $S_B$ is the (2+1)D action governing the bulk fields, $S_{F,\alpha}^{(x)}$ is the (1+1)D defect action supported on the $\alpha$th column, and $S_{F,\alpha}^{(y)}$ is the (1+1)D defect action supported on the $\alpha$th row. We should ensure that the fields corresponding to all other string sectors decouple, and that we can take the continuum limit in a sensible way. We claim that this can be achieved by taking
\begin{align}\label{eqn:option1}
    g_s,l_s,\delta \to 0 \text{ such that } M_{\mathrm{W}}\sim \delta/l_s^2 \to\infty, \text{ and } g_{\mathrm{YM}}^2 \sim g_s/l_s \text{ is fixed and finite.}
\end{align}
Indeed, as explained in \S\ref{sec:branereminders}, we must take $l_s\to 0$ to isolate the light open string states from the rest of the string modes. In order for the coupling in Eq.\ \eqref{eqn:schematicd2d4d4p} to remain finite, we must then also take $g_s\to 0$ in tandem so that $g_{\mathrm{YM}}^2$ is fixed. As described towards the end of \S\ref{sec:branereminders},  the ${^\mathrm{F}}$D4$_\alpha^{(x)}$-${^\mathrm{F}}$D4$_\alpha^{(x)}$ and ${^\mathrm{F}}$D4$_\alpha^{(y)}$-${^\mathrm{F}}$D4$_\alpha^{(y)}$ strings then become free and decouple (because the Yang-Mills coupling on the world-volume of D4-branes goes to zero in the limit of Eq.\ \eqref{eqn:option1}). What was once the gauge group on the world-volumes of the D4-branes then becomes a global symmetry group from the perspective of the D2-branes. Because we took $M_{\mathrm{W}}\to\infty$, the W-bosons corresponding to ${^\mathrm{F}}$D4$_\alpha^{(x)}$-${^\mathrm{F}}$D4$_{\alpha'}^{(x)}$ and ${^\mathrm{F}}$D4$_\alpha^{(y)}$-${^\mathrm{F}}$D4$_{\alpha'}^{(y)}$ strings stay massive, and so that global symmetry group is $\textsl{U}(M)^{2L}$, as opposed to the larger $\textsl{U}(2ML)$ group that we would find if the W-bosons became massless. Finally, the ${^\mathrm{F}}$D4$_\alpha^{(x)}$-${^\mathrm{F}}$D4$_\alpha^{(y)}$ strings are also effectively frozen and decouple because the $x^3,\dots,x^6$ directions are infinite in size, and so their kinetic terms are very large.

So in total, we find that the action in Eq.\ \eqref{eqn:schematicd2d4d4p} completely captures the world-volume theory of the brane web. If we would like to  convert the sum over $\alpha$ to an integral in order to make closer contact with foliated field theories, we can make a field redefinition of all the fields in the defect hypermultiplets, $X' = X/\sqrt{\delta}$. Then, any term quadratic in the defect hypermultiplet fields (including e.g.\ covariant derivative terms which couple the bulk gauge field to the defect fields) comes with a factor of $\delta$ out front which combines with the sum over $\alpha$ to turn into an integral in the direction orthogonal to the leaves of the foliation. However, interaction terms involving three or more fields of the defect hypermultiplets (like Yukawa terms) are suppressed by higher powers of $\delta$, and so the leaves are essentially free, except for interactions which are mediated by the bulk fields.

The reason we have kept $N$, $M$ general in this discussion is to facilitate possible future investigations into this brane system via holography, which requires one to take a large $N$ limit. However, for the purposes of studying the world-volume field theory in its own right, it is most natural and simplest to take $N=M=1$. Although it is very possible to write down an explicit Lagrangian for this theory, e.g.\ by adapting the techniques of \cite{DeWolfe:2001pq}, we will content ourselves, in this section at least, with simply pointing out that the purely bosonic part of the action will look similar, in the case that $N=M=1$, to Eq.\ \eqref{eqn:complexfftnonsusy2p1d}. The difference is that there will be two species of fields for each of the two foliations  corresponding to the two complex scalars of the defect hypermultiplets, i.e.\ the fields  $\Phi^{(x)},\widetilde{\Phi}^{(x)}$ and $\Phi^{(y)},\widetilde{\Phi}^{(y)}$. Moreover, the bulk will include 7 extra scalars $X^I$ in addition to the gauge field.

\subsection{Planar $\normalfont\textsl{U}(1)$ subsystem symmetry in (3+1)D and duality}\label{subsec:planar}

We now carry out a similar construction in one dimension higher, this time introducing dualities into the mix.  We will simplify the discussion by focusing throughout on the case that there is only a single background foliation field, leaving the interesting case of multiple foliation fields to future work. Much of the content of this section can be thought of as a re-interpretation/adaptation of existing techniques to the setting of foliated field theory.

\subsubsection{An analogy: ordinary $\normalfont \textsl{U}(1)$ symmetry in (2+1)D and duality}\label{subsubsec:analogy}
Because planar subsystem symmetries in (3+1)D behave in many ways like ordinary symmetries in (2+1)D, we will start by reminding the reader of some well-known examples of theories that fall into the latter class, the dualities that they enjoy, and how they arise on branes.
We focus on a chain of three increasingly sophisticated dualities, with the punchline being that the last one can be thought of as a parent duality from which the other two (and several others) follow. Moreover this last one can be naturally embedded into string theory and understood as a consequence of the $\textsl{SL}(2,\mathbb{Z})$ S-duality of Type IIB. 

Many good reviews exist on this subject; we recommend \cite{turner2019lectures}, where most of the content covered in this subsection can be found.

\paragraph{Electromagnetic duality}

The simplest example of a dual pair of (2+1)D theories with $\textsl{U}(1)$ global symmetry is free Maxwell theory and the free compact boson; their duality generalizes T--duality of the (1+1)D compact boson. The action of free Maxwell theory is simply 
\begin{align}
    S_{\mathrm{EM}}[a] = -\frac{1}{2e^2}\int f\wedge \star f, \ \ \ \ \ f=da.
\end{align}
This theory possesses a magnetic $\textsl{U}(1)$ global symmetry, whose associated conserved current is simply $j^\mu = \frac{1}{2\pi}\epsilon^{\mu\nu\rho}\partial_\nu a_\rho$, or %
\begin{align}
    j = \frac{1}{2\pi}\star da
\end{align}
in the notation of differential forms. The conservation of this symmetry is essentially due to a Bianchi identity, i.e.\ it is identically conserved without even the requirement to appeal to the equations of motion. It will be useful to couple electrodynamics to a background gauge field $A$ for this magnetic $\textsl{U}(1)$. This can be accomplished by adding a ``BF'' term to the action,
\begin{align}
    S_{\mathrm{EM}}[a;A] =  \int\left(-\frac{1}{2e^2} f\wedge\star f+\frac{1}{2\pi}A\wedge da \right).
\end{align}
Because it is a background field, we do \emph{not} integrate over $A$ in the path integral.

To produce a dual theory, one notices that the action above depends on $a$ only through its field strength $f$. Therefore, instead of integrating over $a$ in the path integral, one is tempted to simply integrate over $f$. This change of variables can be done, however since $f$ is exact, one should impose that $df=0$ as a constraint to make this sensible. To achieve this, one introduces a Lagrange multiplier field $\sigma$ (the ``dual photon''), so that
\begin{align}
\begin{split}
    \mathcal{Z}[A] &= \int\mathcal{D}a\exp i\int\left(-\frac{1}{2e^2} f\wedge\star f+\frac{1}{2\pi}A\wedge da \right) \\
    &= \int \mathcal{D}f \mathcal{D}\sigma \exp i \int \left(  -\frac{1}{2e^2} f\wedge\star f+\frac{1}{2\pi}A\wedge da+\frac{1}{2\pi}\sigma df \right)
\end{split}
\end{align}
where $\sigma$ should be compact (with period $2\pi$) to reflect the quantization of monopole charge. Since the action depends linearly on $\sigma$, integrating over it in the path integral amounts to plugging in its equation of motion. That equation of motion is $df=0$, which means that $f$ can be locally expressed as the exterior derivative of a gauge field $a$, thus landing us back on free Maxwell theory. 

A ubiquitous trick is then to change the order of integration, i.e.\ integrate out $f$ first. Again, since the path integral depends quadratically on $f$, this simply amounts to plugging in its equation of motion. If one does this, one finds
\begin{align}\label{eqn:dualphoton}
\mathcal{Z}[A] = \int\mathcal{D}\sigma \exp\left( i
    S_{\mathrm{DP}}[\sigma;A]\right), \ \ \ \  S_{\mathrm{DP}}[\sigma;A]= \frac{e^2}{8\pi^2}\int (d\sigma-A)\wedge\star(d\sigma -A).
\end{align}
 The magnetic $\textsl{U}(1)$ of the photon is identified with the shift symmetry of the dual photon, and accordingly the background gauge field $A$ is coupled to $\sigma$ through this symmetry, whose corresponding conserved current is simply $j^\mu = -\frac{e^2}{4\pi}\partial^\mu\sigma$, or 
\begin{align}
    j = -\frac{e^2}{4\pi^2}d\sigma.
\end{align}
We emphasize that this duality is an exact duality between free theories, and so although it may appear surprising, it can be demonstrated easily using path integral techniques, as we have seen. 

\paragraph{Bosonization}

As we have stated several times, our ultimate ambition is to realize these kinds of field theories and dualities on branes. This runs into the complication that the brane configurations we consider more naturally give rise to complex scalars, rather than real compact scalars.\footnote{Of course, the electromagnetic duality of the previous subsection famously arises in relating the gauge field on a D2-brane in Type IIA to the embedding coordinate of the dual M2-brane into the extra 11th dimension of M-theory. However, it is not clear how to generalize this fact to the setting of subsystem symmetries.} The way to get around this is that one can embed a real compact scalar $\sigma$ into a complex field $\Phi$ by realizing the former as the phase mode of the latter, $\Phi\sim \rho e^{i\sigma}$. Doing this for the theory of the dual photon leads us to 
\begin{align}
    |D_{-A} \Phi|^2-\mu|\Phi|^2-\lambda |\Phi|^4.
\end{align}
Indeed, if we tune $\mu$ to be negative and $\lambda$ positive, then the $\textsl{U}(1)$ global symmetry of this theory is spontaneously broken, and the dual photon is the Goldstone boson of this spontaneous symmetry breaking. In the IR, the theory goes over to Eq.\ \eqref{eqn:dualphoton}. 

Meanwhile, it is known that this theory is dual to SQED, decorated by a contact term 
\begin{align}
    -\frac{1}{2e^2}f\wedge\star f + \bar\psi  (i\slashed{D}_a-\tilde{\mu})\psi +\frac{1}{2\pi} Ada - \frac{1}{4\pi} AdA
\end{align}
with the negative $\mu$ phase being mapped to the positive $\tilde{\mu}$ phase, and vice versa. (cf.\ e.g.\ \cite{peskin1978mandelstam,dasgupta1981phase} for some of the original papers, and also \cite{Karch:2016sxi,seiberg2016duality} for more recent accounts of similar dualities.) Indeed, when $\tilde{\mu}\gg0$, the fermion is gapped out, leaving just a free Maxwell theory of $a$ in the IR. Thus, this duality is a generalization of the electromagnetic duality we encountered in the previous subsection.

\paragraph{Mirror symmetry from S--duality of Type IIB  string theory}

For our purposes, it will be useful to generalize the dualities of the previous subsections one step further by incorporating supersymmetry. Specifically, we will work with (2+1)D gauge theories with $\mathcal{N}=4$ supersymmetry. The relevant field theory dualities go by the name of \emph{mirror symmetry} (related to, but distinct from mirror symmetry of (1+1)D quantum field theories), and were first discovered in \cite{intriligator1996mirror}. For our purposes, it will be easiest to describe these dualities in the context of brane physics.

\begin{table}[]
    \centering
    \begin{tabular}{c|ccc|cccccc c}
    & $t$ & $x$ & $y$ & $3$ & $4$ & $5$ & $6$ & $7$ & $8$ & $9$ \\\hline
         D3 & \textbf{x} & \textbf{x} & \textbf{x} & & & & \textbf{x}  \\
         D5 & \textbf{x} & \textbf{x} & \textbf{x} & \textbf{x} & \textbf{x} & \textbf{x} & $0$ \\
         NS5$_L$ & \textbf{x} &\textbf{x} &\textbf{x} & &&&$-\epsilon$ &\textbf{x} &\textbf{x} &\textbf{x} \\
         NS5$_R$ & \textbf{x} &\textbf{x} &\textbf{x} & &&&$\epsilon$&\textbf{x} &\textbf{x} &\textbf{x} \\ 
    \end{tabular}
    \caption{A Hanany-Witten brane setup. An \textbf{x} denotes a direction that the brane spans. The numbers $0,\pm \epsilon$ are meant to denote the positions of corresponding branes in the $x^6$ dimension.}
    \label{tab:hananywitten}
\end{table}

\begin{figure}
    \centering
    \begin{tikzpicture}[cross/.style={path picture={ 
  \draw[black]
(path picture bounding box.south east) -- (path picture bounding box.north west) (path picture bounding box.south west) -- (path picture bounding box.north east);
}}]
\draw[->] (-2.2-2,.4) -- (-2.2+.9-2,.4) node[anchor= north] {$x^6$};
\draw[->] (-2.2-2,.4) -- (-2.2-2,1.3) node[anchor= east] {$x^5$};
\draw[thick] (0,.8) -- (3,.8);
\draw[thick,dashed] (1.5,0) -- (1.5,1.6);
\node [draw,fill=white,thick,circle,cross,minimum width=.3 cm] at (3,.8){};
\node [draw,fill=white,thick,circle,cross,minimum width=.3 cm] at (0,.8){};
\node [draw,fill=white,thick,circle,cross,minimum width=.3 cm] at (-3.3-2,.5){};
\node[] at (-3.3-2,1) {$x^7$};
\end{tikzpicture}
    \caption{A visualization of the brane setup in the $x^5$-$x^6$ plane, with the $x^7$ direction being into the page. The horizontal line is the D3 brane. The two crossed circles represent the NS5$_L$ and NS5$_R$-branes. The vertical dashed line is the D5 brane.}
    \label{fig:hananywitten}
\end{figure}
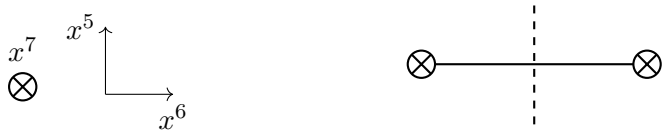

Following \cite{hanany1997type}, we work with webs of D3-branes, D5-branes, and NS5-branes. For illustrative purposes, we consider a simple example in which the world-volumes of these branes are arranged in spacetime as in Table \ref{tab:hananywitten} and Figure \ref{fig:hananywitten}. This configuration preserves $\mathcal{N}=4$ supersymmetry in the (2+1)D sense. Its physics is very similar to the D3/D5 system we briefly described at the beginning of \S\ref{subsec:D2D4D4}. In particular, the role of the D5 brane as before is to contribute a fundamental hypermultiplet which describes the D3-D5 strings. The main new feature is the introduction of the NS5-branes, which serve two purposes. First of all, they truncate the world-volume of the D3-brane in the $x^6$ direction; if we take $\epsilon\to 0$ (i.e.\ if we take the NS5-branes to be very close to each other in the $x^6$ direction), we can think of the resulting theory supported on the world-volume of the D3-brane as a (2+1)D theory, just as in dimensional reduction. Secondly, the NS5-branes impose 1/2-BPS boundary conditions which set some of the fields one would normally encounter on the world-volume of a D3-brane to zero. More specifically, without the NS5-branes present, the D3-brane would support a (2+1)D $\mathcal{N}=8$ vector-multiplet, which decomposes into an $\mathcal{N}=4$ vector-multiplet and an $\mathcal{N}=4$ adjoint hypermultiplet; the NS5-branes project out all of the fields in the hypermultiplet. In total, we are left with a (2+1)D $\mathcal{N}=4$ theory of a vector-multiplet coming from the D3-D3 strings, and a charge 1 hyper-multiplet coming from D3-D5 strings. This theory can be thought of as a supersymmetric variant of quantum electrodynamics in (2+1)D. 

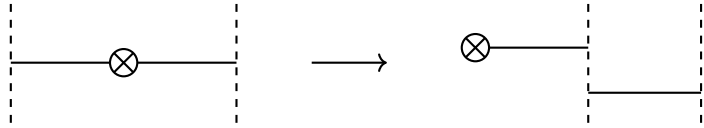
\begin{figure}
    \centering
    \begin{tikzpicture}[cross/.style={path picture={ 
  \draw[black]
(path picture bounding box.south east) -- (path picture bounding box.north west) (path picture bounding box.south west) -- (path picture bounding box.north east);
}}]
\draw[thick] (0,.8) -- (3,.8);
\draw[thick,dashed] (0,0) -- (0,1.6);
\draw[thick,dashed] (3,0) -- (3,1.6);
\node [draw,fill=white,thick,circle,cross,minimum width=.3 cm] at (1.5,.8){};
\draw[->,thick] (4,.8) -- (5,.8);
\end{tikzpicture}
\hspace{.28in}
    \begin{tikzpicture}[cross/.style={path picture={ 
  \draw[black]
(path picture bounding box.south east) -- (path picture bounding box.north west) (path picture bounding box.south west) -- (path picture bounding box.north east);
}}]
\draw[thick] (0,1) -- (1.5,1);
\draw[thick] (1.5,.4) -- (3,.4);
\draw[thick,dashed] (1.5,0) -- (1.5,1.6);
\draw[thick,dashed] (3,0) -- (3,1.6);
\node [draw,fill=white,thick,circle,cross,minimum width=.3 cm] at (0,1){};
\end{tikzpicture}
\caption{On the left, a visualization of the brane setup after S-dualizing. After moving the left-most D5-brane through the NS5-brane, the picture becomes that on the right.}\label{fig:hananywittensdual}
\end{figure}

Now we consider applying S-duality of Type IIB string theory. This duality maps D3-branes to D3-branes and exchanges D5-branes with NS5-branes. Thus, up to a rotation of spacetime, we find that after dualizing we end up with the brane configuration on the left of Figure \ref{fig:hananywittensdual}. The results of \cite{hanany1997type} inform us that it is convenient to move the left-most D5-brane through the NS5-brane before attempting to analyze the brane intersection, whence we end up with the configuration on the right side of Figure \ref{fig:hananywittensdual}. Let us analyze the field theory which arises now on the world-volume of the D3-brane. For the segment of D3-brane which stretches between the NS5-brane and the D5-brane, the theory is essentially trivially gapped in the IR. The reason for this is that the NS5-brane imposes boundary conditions which project out the (2+1)D $\mathcal{N}=4$ adjoint hypermultiplet which would ordinarily appear on the world-volume of a D3, and the D5-brane projects out the (2+1)D $\mathcal{N}=4$ vectormultiplet, so there are simply no fields left. For the segment of D3-brane which is stretched between two D5-branes, just the vectormultiplet is projected out, leaving only the theory of a hypermultiplet. 

Thus in total, by analyzing the dual brane configurations, we learn that SQED$_\mathrm{3}$ is dual to a theory of a hypermultiplet. This is a supersymmetric promotion of the bosonization duality of the previous section. Indeed, one can derive the non-supersymmetric duality from this mirror pair by studying supersymmetry breaking perturbations \cite{tong2000dynamics,kachru2016bosonization,kachru2017nonsupersymmetric}. 

\subsubsection{Non-supersymmetric prototypes}

The philosophy of foliated field theory is that we can promote all of the theories of the previous subsection to theories in (3+1)D with planar subsystem symmetries by foliating space with them, and possibly coupling them to some bulk fields. 

Let us try this with the dual photon, following a protocol similar to that in \S\ref{subsubsec:linearfoliatednonsusy}. To simplify the discussion, we will work in flat Minkowski space with the $z$-direction wrapped into a circle of radius $R$. We also specialize to the case of a single trivial foliation field, $e=dz$, though many of our formulae can be adapted to the more general case. Our action is 
\begin{align}\label{eqn:4dfftdualphoton}
\begin{split}
    S[A,\sigma] &= S_B[dA]+S_F[\sigma,A], \\
    S_B[F] &= \int
    \left(-\frac{1}{2e^2} F\wedge\star F + \frac{\theta}{8\pi^2} F\wedge F\right)\\
    &=\int d^4x\left( -\frac{1}{4e^2}F_{MN}F^{MN} + \frac{\theta}{32\pi^2} \epsilon^{MNPQ} F_{MN}F_{PQ}\right), \\
    S_F[\sigma,A]&= \frac{g^2}{8\pi^2}\int (d\widehat\sigma-\widehat A)\wedge\star (d\widehat\sigma-\widehat A) \\
    &= \frac{g^2}{8\pi^2}\int d^4x (\partial_\mu\sigma-A_\mu) (\partial^\mu\sigma-A^\mu).
\end{split}
\end{align}
In the above, $\sigma$ is a compact, real scalar field, $F=dA$ is the field strength of a (3+1)D $\textsl{U}(1)$ gauge field $A$, and $M,N,\dots$ run over $t,x,y,z$ while $\mu,\nu,\dots$ run over $t,x,y$. Also, generalizing the notation of \S\ref{subsubsec:linearfoliatednonsusy}, we write $\widehat\sigma \equiv \sigma\wedge dz$, and likewise $\widehat{A}\equiv A\wedge dz$. This is the action of bulk electrodynamics coupled to the compact free bosons on the leaves of the foliation through their $\textsl{U}(1)$ global symmetry. As before, this theory has a subsystem symmetry group supported on the leaves of the foliation, and which acts as
\begin{align}
    \sigma(t,x,y,z) \to \sigma(t,x,y,z)+c(z).
\end{align}

This action has two kinds of dualities. First of all, one can dualize each of the bosons on the leaves of the foliation to photons. This is a straightforward generalization of (2+1)D electromagnetic duality, leaving us with 
\begin{align}
\begin{split}
    \widetilde{S}[A,a] &= S_B[dA]+\widetilde{S}_F[a,A] \\
    S_B[F] &= \int \left(-\frac{1}{2e^2} F\wedge\star F + \frac{\theta}{8\pi^2} F\wedge F\right)\\
    &=\int d^4x\left( -\frac{1}{4e^2}F_{MN}F^{MN} + \frac{\theta}{32\pi^2} \epsilon^{MNPQ} F_{MN}F_{PQ}\right) \\
    \widetilde{S}_F[a,A]&= \int\left(-\frac{1}{2g^2}  d\widehat a \wedge\star d\widehat a+\frac{1}{2\pi} A\wedge d\widehat{a}\right) \\
    &= \int d^4x \left(-\frac{1}{4g^2}f_{\mu\nu}f^{\mu\nu}+\frac{1}{2\pi}\epsilon^{\mu\nu\rho}A_\mu\partial_\nu a_\rho \right).
\end{split}
\end{align}
On the other hand, we can also attempt to perform S-duality of the bulk (3+1)D electromagnetism. This duality requires more care. Our perspective is that the theory Eq.\ \eqref{eqn:4dfftdualphoton} is somewhat ill-defined, as it features bulk electromagnetism coupled to a continuum of defect theories. In particular, the $\sigma$ field has no kinetic term in the $z$-direction, and so arbitrarily short distance fluctuations are unsuppressed. To make it better-defined, we should regularize it. Our choice of regularization is natural from the point of view of D-branes: we discretize the foliation in the $z$-direction. That is, we demote the continuous $z$-coordinate of the foliation field $\sigma(t,x,y,z)$ to a discrete index, $\sigma^{I}(t,x,y)\sim \sigma(t,x,y,z=I\delta)$, with $\delta$ the regularization parameter, and $I$ an integer running from $1$ to $L$, the regularized number of leaves. We think of the field $\sigma^{I}$ as living on the leaf $\mathcal{L}_I = \{(t,x,y,z=I\delta)\subseteq \mathbb{R}^{1,2}\times S^1\}$. The foliation action then can be approximated as
\begin{align}
    S_F[\sigma,A]\sim \sum_{I=1}^{L} S_F^{I}[\sigma^{I},A] \equiv \delta\frac{g^2}{8\pi^2} \sum_{I=1}^L \int_{\mathcal{L}_I} d^3x(\partial_\mu\sigma^{I} - A_\mu)(\partial^\mu \sigma^{I} - A^\mu).
\end{align}
In tandem with this, it will facilitate our analysis if we split the (3+1)D bulk gauge field $A$ into a sequence of gauge fields that are defined in the bulk regions between neighboring leaves. More precisely, we split $A$ into gauge fields $A^{I}$ such that $A^{I}$ is defined on $B_I$, the region between the leaves $\mathcal{L}_I$ and $\mathcal{L}_{I+1}$. We then stitch these together by integrating in a one-form Lagrange multiplier $c^{I+1}$ that sets $A^{I} = A^{I+1}$ (up to a gauge transformation) at the leaf $\mathcal{L}_{I+1}$. In total, the action now looks like
\begin{align}
\begin{split}
   \mathcal{Z} &= \int \mathcal{D}A\mathcal{D}\sigma e^{iS[A,\sigma]} \\
   &= \int \prod_I \mathcal{D} A^{I}\mathcal{D}\sigma^{I}\mathcal{D} c^{I}\exp i\sum_I \Bigg(  \int_{B_I}\left(-\frac{1}{2e^2}F^{I}\wedge \star F^{I}+\frac{\theta}{8\pi^2} F^{I}\wedge F^{I} \right) \\
   & \hspace{1in} -\frac{1}{2\pi} \int_{\mathcal{L}_I} c^{I}\wedge d(A^{I}-A^{I-1})+S_F^{I}[\sigma^{I},A^{I}]\Bigg)
\end{split}
\end{align}
Now we can attempt to apply standard path integral techniques for deriving S-duality of (3+1)D electromagnetism, following \cite{gaiotto2009s,gaiotto2009supersymmetric,seiberg2016duality}. We start by integrating over $F^{I}$ and $A^{I}$ as separate fields in the path integral. This is acceptable provided we supplement the theory with the constraint that $F^{I}=dA^{I}$ in each bulk region $B_I$, which can be achieved with a two-form Lagrange multiplier $\widetilde{F}^{I}$. That is, 
\begin{align}
\begin{split}
    \mathcal{Z} &= \int \prod_I \mathcal{D} A^{I}\mathcal{D}F^{I}\mathcal{D}\widetilde{F}^{I}\mathcal{D}\sigma^{I}\mathcal{D} c^{I} \\
    & \exp i\sum_I \Bigg(  \int_{B_I}\left(-\frac{1}{2e^2}F^{I}\wedge \star F^{I}+\frac{\theta}{8\pi^2} F^{I}\wedge F^{I} \right)+\frac{1}{2\pi} \int_{B_I} \widetilde{F}^{I}\wedge (F^{I}-dA^{I}) \\
   & \hspace{1in} -\frac{1}{2\pi} \int_{\mathcal{L}_I} c^{I}\wedge d(A^{I}-A^{I-1})+S_F^{I}[\sigma^{I},A^{I}]\Bigg).
\end{split}
\end{align}
In this expression, we think of $F^{I}$ and $A^{I}$ as unrelated; it is only after integrating out the Lagrange multiplier $\widetilde{F}^{I}$ that one encounters a delta function setting $F^{I}$ equal to the field strength of $A^{I}$. 

In the next step, we can integrate out $F^{I}$, as the action depends quadratically on it. Doing this leads to 
\begin{align}\label{eqn:intermediate}
\begin{split}
    \mathcal{Z} &= \int \prod_I \mathcal{D} A^{I}\mathcal{D}\widetilde{F}^{I}\mathcal{D}\sigma^{I}\mathcal{D} c^{I}\exp i\sum_I \Bigg(  \int_{B_I}\left(-\frac{1}{2\tilde{e}^2}\widetilde{F}^{I}\wedge \star \widetilde{F}^{I}+\frac{\tilde\theta}{8\pi^2} \widetilde{F}^{I}\wedge \widetilde{F}^{I} \right) \\
   & \hspace{.3in} -\frac{1}{2\pi} \int_{B_I} \widetilde{F}^{I}\wedge dA^{I}-\frac{1}{2\pi} \int_{\mathcal{L}_I} c^{I}\wedge d(A^{I}-A^{I-1})+S_F^{I}[\sigma^{I},A^{I}]\Bigg)
\end{split}
\end{align}
where in the above, $\tilde{e}$ and $\tilde{\theta}$ are defined by 
\begin{align}\label{eqn:dualcouplings}
\tilde{\theta}/2\pi+ 2\pi i/\tilde{e}^2 = \frac{-1}{\theta/2\pi + 2\pi i/e^2}. 
\end{align}
We can then integrate out the bulk part of $A^{I}$; by this, we mean that we integrate out the part of the field that is defined  away from the leaves which bound it. The first term on the second line in Eq.\ \eqref{eqn:intermediate} then acts as a Lagrange multiplier which sets $d\widetilde{F}^{I} = 0$ away from the leaves so that we can think of $\widetilde{F}^{I}$ as the field strength of a gauge field $\widetilde{A}^{I}$ in these bulk regions. With the bulk $A^{I}$ integrated out, this term then reduces to a BF-type term on the bounding leaves, 
\begin{align}
\begin{split}
-\frac{1}{2\pi}\int_{B_I}\widetilde{F}^{I}\wedge dA^{I} &= -\frac{1}{2\pi} \int_{B_I} d(\widetilde{A}^{I} \wedge dA^{I} ) \\
&= -\frac{1}{2\pi}\int_{\mathcal{L}_{I+1}}\widetilde{A}^{I}\wedge dA^{I}  + \frac{1}{2\pi} \int_{\mathcal{L}_I} \widetilde{A}^{I}\wedge dA^{I}.
\end{split}
\end{align}
Thus, all that remains of the original gauge fields $A^{I}$ are their restrictions to the leaves which surround them. We use the notation $b^{I}_-= A^{I}\vert_{\mathcal{L}_I}$  and $b^{I}_+=A^{I}\vert_{\mathcal{L}_{I+1}}$ to denote these restrictions. In terms of these variables, we are left with
\begin{align}
\begin{split}
    \mathcal{Z} &=\int \prod_I \mathcal{D}b_\pm^{I}\mathcal{D}c^{I}\mathcal{D}\widetilde{A}^{I}\mathcal{D}\sigma^{I} \\
    & \hspace{.3in} \exp i \sum_{I} \Bigg( \int_{B_{I}} \left(-\frac{1}{2\tilde{e}^2}\widetilde{F}^{I}\wedge \star \widetilde{F}^{I} + \frac{\tilde{\theta}}{8\pi^2} \widetilde{F}^{I} \wedge \widetilde{F}^{I} \right)+S_F^{I}[\sigma^{I},b_-^{I}] \\
    &\hspace{.8in} -\frac{1}{2\pi} \int_{\mathcal{L}_I} \left( c^{I}\wedge d(b_-^{I}-b_+^{I-1})+ \widetilde{A}^{I}db_-^{I}-\widetilde{A}^{I-1}db_+^{I-1} \right)\Bigg) .
\end{split}
\end{align}
Finally, we can integrate out $b_+^{I-1}$ on each leaf. It appears as a Lagrange multiplier in the path integral which sets $c^{I} = -\widetilde{A}^{I-1}$. Calling then $b^{I}=b_-^{I}$ the remaining (2+1)D gauge field on the leaf $\mathcal{L}_I$ gives us
\begin{align}\label{eqn:dualwithoutcontinuum}
\begin{split}
     \mathcal{Z} &=\int \prod_I \mathcal{D}b^{I}\mathcal{D}\widetilde{A}^{I}\mathcal{D}\sigma^{I} \\
    & \hspace{.3in} \exp i \Bigg(\sum_{I} \int_{B_{I}} \left(-\frac{1}{2\tilde{e}^2}\widetilde{F}^{I}\wedge \star \widetilde{F}^{I} + \frac{\tilde{\theta}}{8\pi^2} \widetilde{F}^{I} \wedge \widetilde{F}^{I} \right) \\
    &\hspace{.8in} -\frac{1}{2\pi} \int_{\mathcal{L}_I} b^{I}\wedge d\left(\widetilde{A}^{I}-\widetilde{A}^{I-1} \right)+S_F^{I}[\sigma^{I},b^{I}]\Bigg).
\end{split}
\end{align}
We note that nothing in this calculation relied on $S_F^{I}$ being the action of a compact scalar; we could equally well take it to be the action of any (2+1)D theory with a non-anomalous $\textsl{U}(1)$ symmetry.

The result of performing S-duality so far can be summarized as this. First, the original gauge field $A$ which permeated through all of spacetime dualizes to a sequence of gauge fields $\widetilde{A}^{I}$ which are defined in the bulk regions $B_I$ between the leaves. In their own region, they enter with the standard Maxwell action, but with new couplings defined by $\tilde{e}$ and $\tilde{\theta}$ as defined by Eq.\ \eqref{eqn:dualcouplings}. There is \emph{no} boundary condition which sets $\widetilde{A}^{I}$ equal to $\widetilde{A}^{I+1}$ on the leaf which separates them. Such a structure already is reminiscent of the topological defect network paradigm advocated for in \cite{aasen2020topological}, where one defines a network of theories supported on various ``$k$-strata''; the major difference here is that our theories can be gapless.  Next, one introduces a new (2+1)D gauge field $b^{I}$ on each leaf of the foliation. This gauge field couples both to the $\widetilde{A}^{I}$ through a ``bifundamental BF term'', and also to the action $S_F^{I}$ that was originally on the leaves. 

Now after we have dualized, we would like to restore the continuum limit, $\delta\to 0$ and $L\to\infty$. The main point is that this causes each bulk region $B_I$ to become arbitrarily small in the $z$-direction; thus, we can approximate the action in these regions by dimensional reduction. Maxwell theory in (3+1)D reduces when placed on a small interval $[0,\delta]$ to a Maxwell-dilaton theory of the form 
\begin{align}
\begin{split}
    &\int_{\mathbb{R}^{1,2}\times [0,\delta]} \left(-\frac{1}{2e^2}F\wedge\star F + \frac{\theta}{8\pi^2}F\wedge F\right) \\
    & \ \ \ \ \to S_{\mathrm{MD}}[a,\phi] \equiv \delta\int_{\mathbb{R}^{1,2}}d^3x\left(-\frac{1}{4e^2}f_{\mu\nu}f^{\mu\nu} -\frac{1}{2e^2}\partial_\mu\phi\partial^\mu\phi + \frac{\theta}{16\pi^2}\epsilon^{\mu\nu\rho}\partial_\mu\phi f_{\nu\rho}\right)
\end{split}
\end{align}
where we have defined $a_\mu = A_\mu$ for $\mu=t,x,y$, and also $f=da$ and $\phi = A_3$. Thus, plugging this into Eq.\ \eqref{eqn:dualwithoutcontinuum}, we find in total that the dual of Eq.\ \eqref{eqn:4dfftdualphoton} is 
\begin{align}\label{eqn:fullydualized}
\begin{split}
    &\widetilde{S}[\widetilde{a}^{I},\widetilde{\phi}^{I},b^{I},\sigma^{I}]  \\
    & \ \ \ \ \ \ \ = \sum_I\left( S_{\mathrm{MD}}[\widetilde{a}^{I},\widetilde{\phi}^{I}]-\frac{1}{2\pi} \int (\widetilde{a}^{I}-\widetilde{a}^{I-1})\wedge db^{I}+S_F^{I}[\sigma^{I},b^{I}]\right).
\end{split}
\end{align}
Every ingredient in this action is (2+1)-dimensional. What used to be the $z$-direction has disappeared into the discrete integer $I$, where we roughly think of $z\sim I\delta$. Note that both the first and third terms are multiplied by $\delta$. It is extremely tempting then to multiply the second term by $\delta/\delta$ and write the $\delta\to 0$ continuum limit as 
\begin{align}
\begin{split}
    \widetilde{S}[\widetilde{a},\widetilde{\phi},b,\sigma]&\sim \int_{\mathbb{R}^{1,3}} \Big( -\frac{1}{4\tilde e^2}\widetilde{f}_{\mu\nu}\widetilde{f}^{\mu\nu} -\frac{1}{2\tilde e^2}\partial_\mu\widetilde\phi\partial^\mu\widetilde\phi + \frac{\tilde\theta}{16\pi^2}\epsilon^{\mu\nu\rho}\partial_\mu\widetilde\phi \widetilde f_{\nu\rho} \\
    &  \hspace{1in} -\frac{1}{2\pi}\epsilon^{\mu\nu\rho}\partial_z\widetilde{a}_\mu \partial_\nu b_\rho +\frac{g^2}{8\pi^2} (\partial_\mu\sigma -b_\mu)(\partial^\mu\sigma-b^\mu)  \Big)
\end{split}
\end{align}
where we have traded out the index $I$ for a spacetime coordinate $z$. We emphasize that in the above action, $\mu,\nu,\rho$ run over $t,x,y$ simply because the background foliation field was specialized to $e=dz$. We leave the generalization of this calculation to more general foliations to future work. 

We note in passing the impressionistic similarity of Eq.\ \eqref{eqn:fullydualized} to the idea of deconstruction \cite{arkani2001constructing}, where one takes a (3+1)D gauge theory which is described by a long quiver diagram and attempts to interpret the direction of the quiver as an emergent dimension of a (4+1)D Lorentz invariant theory. We will encounter a similar construction in \S\ref{sec:ics} when we discuss infinite-component Chern-Simons theory.

To summarize, we have described two different dualities of the foliated field theory Eq.\ \eqref{eqn:4dfftdualphoton}, one in which the leaves of the foliation are dualized, and the other in which the bulk is dualized. In the next section, we will see that S-duality of Type IIB string theory performs both of these kinds of dualities at the same time for the analogous supersymmetric foliated field theories.

\subsubsection{A supersymmetric Type IIB brane model and S--Duality}

To conclude, let us describe how the discussion of the previous section can be embedded into string theory. See also \cite{Aitken:2018joz} where a similar setup has appeared, but with a different interpretation.

\begin{table}[]
    \centering
    \begin{tabular}{c|ccccccccc c}
    & $t$ & $x$ & $y$ & $3$ & $4$ & $5$ & $z$ & $7$ & $8$ & $9$ \\\hline
         ${^\mathrm{B}}$D3 & \textbf{x} & \textbf{x} & \textbf{x} & & & & \textbf{x}  \\
         ${^\mathrm{F}}$D5$_\alpha$ & \textbf{x} & \textbf{x} & \textbf{x} & \textbf{x} & \textbf{x} & \textbf{x} & ::
    \end{tabular}
    \caption{A foliated Hanany-Witten brane setup. An \textbf{x} denotes a direction that the brane spans, and :: indicates that the D5-branes form a lattice with spacing $a$ in the $z$ direction.}
    \label{tab:foliatedhananywitten}
\end{table}

\begin{figure}
    \centering
    \begin{tikzpicture}
     \draw[thick] (0cm,0cm) circle(2cm);
     \foreach \x in {0,30,...,360} {
                \draw[dashed,thick] (0cm,0cm) -- (\x:2.5cm);
        }
        \filldraw[white] (0cm,0cm) circle(1.5cm);
    \end{tikzpicture}
    \hspace{.6in}
        \begin{tikzpicture}[cross/.style={path picture={ 
  \draw[black]
(path picture bounding box.south east) -- (path picture bounding box.north west) (path picture bounding box.south west) -- (path picture bounding box.north east);
}}]
\filldraw[white] (0cm,0cm) circle(2.5cm);
     \draw[thick] (0cm,0cm) circle(2cm);
     \foreach \x in {0,30,...,360} {
        \node[draw,fill=white,thick,circle,cross,minimum width=.3 cm] at (\x:2cm){};
        }
    \end{tikzpicture}
    \caption{On the left, the foliated Hanany-Witten setup. On the right, its S-dual. The D3-brane wraps a circle in the $x^6$ direction in both duality frames. The dashed lines are the D5$_\alpha$-branes, and the crossed circles are the NS5$_\alpha$-branes.}
    \label{fig:SdualHananyWitten}
\end{figure}
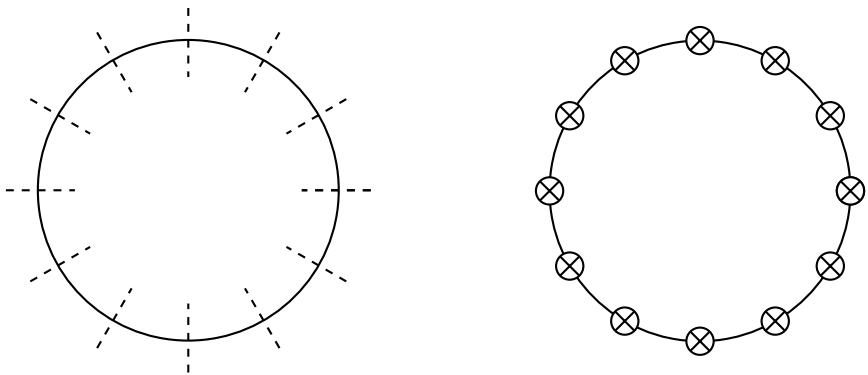

The idea is straightforward. Consider wrapping the $x^6$ direction (which we also call $z$) into a circle of radius $R$. We generalize the situation at the end of \S\ref{subsubsec:analogy} as follows. This time, instead of bounding the world-volume of the D3-brane in the $z$-direction using NS5-branes, we wrap the D3-brane entirely around the circle. Furthermore, we consider a stack of evenly spaced D5-branes in the $z$ direction, with lattice spacing $\delta$. The branes are laid out as in Table \ref{tab:foliatedhananywitten} and Figure \ref{fig:SdualHananyWitten}.

The D3-D3 strings contribute the field content of (3+1)D $\textsl{U}(1)$ $\mathcal{N}=4$ super Yang-Mills. These fields are playing the same role as the bulk Maxwell theory in Eq.\ \eqref{eqn:4dfftdualphoton}; we therefore label this brane as ${^\mathrm{B}}$D3 to emphasize its role in contributing the bulk fields of the foliated field theory. On the other hand, the D3-D5 strings contribute a defect hypermultiplet which is localized on the codimension-1 intersection of the D3-brane and the D5-brane. Hence, the D5-branes are responsible for contributing the fields which live on the leaves of the foliation. We label these branes ${^\mathrm{F}}$D5$_\alpha$, with $\alpha =1,\dots, L$ running over the leaves. The ${^\mathrm{F}}$D5$_\alpha$-${^\mathrm{F}}$D5$_{\alpha'}$ strings decouple, as we've argued previously. In total, we indeed find a discretized and supersymmetric version of the foliated field theory in Eq.\ \eqref{eqn:4dfftdualphoton}. The analysis of the decoupling and continuum limit are similar to the one performed in \S\ref{subsec:D2D4D4}. 

Now we apply S-duality (along with a suitable rotation of space). The ${^\mathrm{B}}$D3-brane is mapped to itself, and the ${^\mathrm{F}}$D5$_\alpha$-branes are exchanged for NS5$_\alpha$-branes. These $L$ NS5$_\alpha$-branes split the ${^\mathrm{B}}$D3-brane into $L$ different segments. Assuming we take the NS5$_\alpha$-branes to be closely spaced, each ${^\mathrm{B}}$D3-brane segment can be thought of as a (2+1)D theory, and the ${^\mathrm{B}}$D3-${^\mathrm{B}}$D3 strings which stretch from one segment to the \emph{same} segment contribute a (2+1)D $\mathcal{N}=4$ vector-multiplet (with the $\mathcal{N}=4$ hypermultiplet projected out by the NS5$_{\alpha}$-branes). On the other hand, the ${^\mathrm{B}}$D3-${^\mathrm{B}}$D3 strings which stretch from one segment to its neighboring segment, crossing an NS5$_\alpha$-brane, contribute a (2+1)D $\mathcal{N}=4$ hypermultiplet which is localized near the NS5$_\alpha$-brane, and which is charged as a bifundamental with respect to the gauge fields living on the two ${^\mathrm{B}}$D3-brane segments it stretches between. Thus, the theory we obtain is a $\textsl{U}(1)^L$ circular quiver gauge theory with $L$ bifundamental hypermultiplets. 

\begin{table}[]
    \centering
    \begin{tabular}{c|ccc|cccccc c}
    & $t$ & $x$ & $y$ & $3$ & $4$ & $5$ & $6$ & $7$ & $8$ & $9$ \\\hline
         ${^\mathrm{B}}$D3 & \textbf{x} & \textbf{x} & \textbf{x} & & & & \textbf{x}  \\
         NS5$_\alpha$ & \textbf{x} &\textbf{x} &\textbf{x} & &&& :: &\textbf{x} &\textbf{x} &\textbf{x} \\
    \end{tabular}
    \caption{The S-dual of the foliated Hanany-Witten brane setup.}
    \label{tab:foliatedhananywittenSdual}
\end{table}

We can now justify our earlier statement that S-duality of Type IIB string theory is performing the composition of both kinds of duality of the foliated field theory (i.e.\ the ``leaf-wise'' duality and the ``bulk'' duality). Let us sketch purely field theoretically what the dualities do, and see that it reproduces what happens on branes.  First, note that if we were to just perform the bulk duality on the supersymmetric foliated-field theory, we would end up with a theory of the form of Eq.\ \eqref{eqn:fullydualized}. The fields on the leaves, which were $\sigma^{I}$ and $b^{I}$ in the non-supersymmetric prototype, are now the original hypermultiplet, and also a new vectormultiplet, on each leaf. That is, each leaf now supports an SQED$_3$ theory. The bulk fields, which were $\widetilde{a}^{I}$ and  $\widetilde{\phi}^{I}$ in the non-supersymmetric prototype, are now bulk vector-multiplets. The global $\textsl{U}(1)$ symmetry of the SQED$_3$ theory on each leaf is coupled as a bifundamental to the bulk vector-multiplets which lie on either side of it. Next, we perform the leaf-wise duality. As we described in \S\ref{subsubsec:analogy}, this is simply mirror symmetry which relates SQED$_3$ to a theory of a hypermultiplet. Thus in total, the composition of the bulk duality and the leaf-wise duality yields a circular quiver of $L$ vector-multiplets coupled to $L$ bifundamental hypermultiplets. This is precisely what S-duality of Type IIB string theory produces.

\section{Exotic field theories}\label{sec:eft}

In this section, our focus is on understanding the kinds of field theories whose study was systematically developed in \cite{seiberg0,seibergshao1,seibergshao2,seibergshao3,seibergshao4,seibergshao5,seibergshao6,seibergshao7,Gorantla:2021bda}. In contrast to FFTs, these field theories  are not coupled to background foliation fields (at least at this stage in their development), nor does their description manifestly require consideration of lower-dimensional defect theories. Instead, the focus is on exotic symmetries, like \emph{subsystem symmetries} which are allowed to transform the fields in a spatially dependent manner. One constructs Lagrangians for such theories in the usual way by enumerating all relevant and marginal terms which are consistent with the symmetry. 

To distinguish this approach from the foliated field theory approach, we refer to such models as \emph{exotic field theories}, borrowing the adjective used to describe their symmetry groups in e.g.\ \cite{seibergshao1}. However, we once again hasten to emphasize that foliated field theories and exotic field theories are thought in many cases to be describing the same underlying physics; in such cases, the difference between them amounts to a difference in presentation. 

In this work, our focus will be on understanding examples with $\textsl{U}(1)$ global symmetries supported at every point in space \cite{seiberg0}, though we expect the techniques we highlight here will generalize to other kinds of subsystem symmetries as well.

\subsection{Non-supersymmetric prototypes}\label{subsec:oneformnonsusy}

The discussion in this section follows \cite{seiberg0} closely. 

As just stated, we will be considering ``point-like" subsystem symmetries:  global symmetries which are allowed to transform the fields independently at each point of space. 
A prototypical example is a theory of a complex scalar $\Phi$ which is invariant under  phase rotations of the form 
\begin{align}\label{eq:s0shift}
    \Phi(t,\vec{x}) \to e^{ic(\vec{x})}\Phi(t,\vec{x})
\end{align}
where $c(\vec{x})$ is an arbitrary function of the spatial coordinates.
The leading order terms consistent with such a symmetry are \cite{seiberg0,pretko:matter}
\begin{align}\label{eqn:pointlikematter}
\mathcal{L}_0 = i\Phi^\dagger\partial_t\Phi - s\partial_i(\Phi^\dagger\Phi)\partial^i(\Phi^\dagger\Phi) - \lambda (\Phi^\dagger\Phi)^2+\cdots
\end{align}
where, as standard in non-relativistic field theories, we have removed a mass term $M^2|\Phi|^2$ by performing a field redefinition of $\Phi$ by a time-dependent phase and then rescaling. It is also perfectly sensible to think of $\Phi$ as a Grassmann field/spinless non-relativistic fermion\footnote{The term proportional to $\lambda$ would then vanish unless one introduced additional species of fermions and formed quartic couplings between them.} in the above action, but we will continue to take $\Phi$ to be a boson for now.

 The symmetry in Eq.\ \eqref{eq:s0shift} is rather large, and correspondingly there are many  charges. 
 Indeed, the Noether current $\mathcal{J^\mu}$ of this symmetry is  given by
\begin{align}\label{eq:s0current}
    \mathcal{J}_t = \Phi^\dagger\Phi + \cdots, \ \ \ \ \ \mathcal{J}^i = 0
\end{align}
and its conservation tells us that $\partial_\mu \mathcal{J}^\mu = \partial_t\mathcal{J}_t = 0$. Thus, $\Phi^\dagger\Phi$ is conserved at every point in space independently, and so we have infinitely many conserved charges.

In passing, we make a simple observation about power-counting in this theory.  For definiteness, we work in 2+1 space-time dimensions.  The invariance of the kinetic and gradient terms under scaling tells us that if $[\dots]$ denotes the mass dimension of a given quantity, then
\begin{gather}
\begin{split}
2[x] + [t] - [t] + 2[\Phi] = 0\\
2[x] + [t] - 2[x] + 4[\Phi] = 0.
\end{split}
\end{gather}
The resulting system then has
\begin{align}
    [x] = -[\Phi]
\end{align}
and
\begin{align}
    [t] = -4[\Phi],
\end{align}
which means that there is a $z=4$ scale invariance under which
\begin{align}
    t \to b^4 t, ~~x \to b x,~~\Phi \to b^{-1} \Phi.
\end{align}
In this scaling, the $(\Phi^\dagger\Phi)^2$ coupling is relevant.  Higher order terms allowed by the symmetry would be marginal
or irrelevant.

Before moving on to our brane configuration, we offer another perspective on the Lagrangian in Eq.\ \eqref{eqn:pointlikematter} which will be useful to have in the back of one's mind. We would like to think of the field $\Phi(t,x,y)$ as arising from taking the continuum limit of a square lattice of quantum mechanical defects $\Phi_\alpha(t)$ with spacing $\delta$, 
\begin{align}
    S_0\sim \sum_{\alpha}\int dt\left( \partial_t\Phi_\alpha^\dagger \partial_t \Phi_\alpha-M^2 \Phi_\alpha^\dagger \Phi_\alpha - \lambda (\Phi^\dagger_\alpha\Phi_\alpha)^2 \right).
\end{align}
In order to produce interactions between defects on neighboring sites, we introduce a (3+1)D gauge field which couples them together via an \emph{instantaneous} Coulomb interaction of strength $g$,
\begin{align}\label{eqn:instantaneouscoulomb}
    S \sim \frac{1}{2g^2}\int dt d^3x (\partial_i A_0)^2 + \sum_\alpha \int dt \left((D_t\Phi_\alpha)^\dagger D_t\Phi_\alpha -M^2 \Phi^\dagger_\alpha\Phi_\alpha - \lambda (\Phi^\dagger_\alpha\Phi_\alpha)^2  \right)
\end{align}
where above, $D_t\Phi_\alpha = (\partial_t-iA_0)\Phi_\alpha$. We can think of obtaining this action by coupling the defect lattice to ordinary relativistic $\textsl{U}(1)$ gauge theory, restoring the speed of light $c$ in all equations, and taking $c\to \infty$. 
Of course, there is no gauge field in Eq.\ \eqref{eqn:pointlikematter}, so to bridge this gap, we will take our continuum limit, $\delta\to 0$, in tandem with taking $g\to 0$ in such a way that the gauge field decouples from the physics, but the interactions it induces nonetheless survive.

To see that this is possible, imagine performing the integral over $A_0$ in the path integral. The structure of the resulting effective action at tree level is\footnote{Some of these coefficients have IR divergences that we can regulate by making space a torus, and some have unimportant UV divergences.}
\begin{align}
\begin{split}
    S_{\mathrm{eff}} &\sim \sum_\alpha\int dt\Big(  i\Phi_\alpha^\dagger\partial_t \Phi_\alpha  -(\lambda+c_1\frac{g^2}{\delta}+\cdots) (\Phi^\dagger_\alpha \Phi_\alpha)^2 \\
    &\hspace{1in}+ (c_2\frac{g^2}{\delta}+\cdots)\Phi^\dagger_\alpha \Phi_\alpha \Delta_i \Delta^i \Phi^\dagger_\alpha \Phi_\alpha + \cdots  \Big)
\end{split}
\end{align}
where $\Delta^i\Delta^i \Phi^\dagger_\alpha \Phi_\alpha$ refers to a finite difference representation of the second derivative, and once again, we have performed a field redefinition by a time-depenent phase and a field rescaling in order to eliminate the mass term.
One may be surprised that we have obtained a theory which is local in time after having integrated out a massless field, however this is why we chose an instantaneous Coulomb interaction. We may then take a continuum limit by defining an effective lattice spacing $\ell$ and rescaling the fields $\Phi_\alpha \to \ell \Phi_\alpha$ so that the action becomes
\begin{align}
\begin{split}
    S_{\mathrm{eff}} &\sim \ell^2\sum_\alpha\int dt\Big(  i\Phi_\alpha^\dagger\partial_t \Phi_\alpha  -\ell^2(\lambda+c_1\frac{g^2}{\delta}+\cdots) (\Phi^\dagger_\alpha \Phi_\alpha)^2 \\
    &\hspace{1in}+ \ell^4(c_2\frac{g^2}{\delta}+\cdots)\Phi^\dagger_\alpha \Phi_\alpha \frac{\Delta_i \Delta^i \Phi^\dagger_\alpha \Phi_\alpha}{\ell^2} + \cdots  \Big)
\end{split}
\end{align}
Thus, if we define $\ell^4\sim \delta/g^2$, and take $\delta,g\to 0$ in such a way that $\ell \to 0$ as well, then in the continuum this goes over to the Lagrangian in Eq.\ \eqref{eqn:pointlikematter}. The various higher derivative terms are suppressed by powers of $\delta/g^2$, and so we recover a local theory. In the strict continuum limit, the quartic terms without spatial derivatives become strong (as befits a relevant coupling) unless one simultaneously tunes $\lambda$ with $\delta$ and $g$ to produce a cancellation.

\subsection{A supersymmetric Type IIB brane model}\label{subsec:pointlikesusy}

\begin{table}
\begin{center}
\begin{tabular}{c|cccc|cccccc}
&$t$&$x$&$y$&$z$&$4$&$5$&$6$&$7$&$8$&$9$ \cr
\hline
$\mathrm{D}3$ & \textbf{x}&\textbf{x}&\textbf{x}&\textbf{x}&&&&&& \\
$\mathrm{D}5_\alpha$ & \textbf{x}& :: & :: &&\textbf{x}&\textbf{x}&\textbf{x}&\textbf{x}&\textbf{x}& $\epsilon$ \\
\end{tabular}
\caption{A diagram encoding the layout of our branes in spacetime. An \textbf{x} denotes a direction along which the branes extend, and :: a direction along which they are localized but form an evenly spaced lattice with spacing $\delta$. The entry with an $\epsilon$ in it emphasizes that we place the D5$_\alpha$-branes at $x^9=\epsilon$, for $\epsilon$ a small number, in order to give the D3-D5$_\alpha$ strings a light mass. If an entry is empty, the brane is at the origin of the corresponding dimension, i.e.\ at $x^i=0$. }\label{fig:D2D2p}
\end{center}
\end{table}

We now attempt to find a brane realization of (a supersymmetric and fermionic version of) the matter theory  described in the previous subsection, Eq.\ \eqref{eqn:pointlikematter}. 
 Various closely related brane models \cite{Jensen:2011su,kachru2010holographic,kachru2011adventures} have been studied e.g.\ in relation to local quantum criticality \cite{si2001locally,gegenwart2008quantum}.

We work in Type IIB string theory on $\mathbb{R}^{1,7}\times S^1_x\times S^1_y$, where both of the circle factors have radius $R$ and are coordinatized by $x\cong x+2\pi R$ and $y\cong y+2\pi R$, respectively. We consider a stack of $N$ D3-branes
extended along the $(x^0,x^1,x^2,x^3) \equiv (t,x,y,z)$ dimensions, and localized at $x^i = 0$ for $i > 3$. The D3-D3 strings contribute a (3+1)D $\mathcal{N}=4$, $\textsl{U}(N)$ super Yang-Mills theory localized on the world-volume of the D3-branes. The bosonic content of this theory is a gauge field and $6$ scalars in the adjoint representation, as in Eq.\ \eqref{eqn:bosonicSYM}. If we take $N=1$, this is an Abelian gauge theory, and, once we introduce matter, this gauge multiplet will furnish a Coulomb interaction as in Eq.\ \eqref{eqn:instantaneouscoulomb}, with coupling constant given as $g_{\mathrm{YM}}\sim g_s$. By restricting our attention to velocities far below the speed of light, we can think of this interaction as instantaneous in the sense described in the previous subsection. We will ultimately take a limit which decouples the gauge multiplet, leaving us with 
a low energy action analogous to Eq.\ \eqref{eqn:pointlikematter}.  

The matter degrees of freedom will be realized by adding single probe D5$_\alpha$-branes extended along the $t, x^{4},\dots, x^8$ dimensions, and which intersect the $x$-$y$ plane along a square lattice with spacing $\delta$ whose sites we label with $\alpha$. In contrast with our earlier 4ND D3/D5 intersections, this brane intersection is 8ND and therefore the D3-D5$_\alpha$ strings lead to purely fermionic supersymmetric multiplets in (0+1)D (once one eliminates auxilliary bosonic fields). We denote the complex fermions in these multiplets $\chi_{\alpha,b}(t)$. We place these D5$_\alpha$-branes at $x^9 = \epsilon$ to give them a mass $M \sim \epsilon/l_s^2$. 

\begin{figure}
    \centering
\begin{tikzpicture}
\fill[blue!30] (-.9,-.9) rectangle (1.6*3+.9,1.6*3+.9);
\foreach \x in {1,...,4}
    \foreach \y in {1,...,4} 
    {\filldraw (1.6*\x-1.6,1.6*\y-1.6) circle (3pt);
    \node[] at (1.6*\x-1.6+.3,1.6*\y-1.6+.3) {D5$_{(\x,\y)}$};};
    \node[] at (0-.5,-.5) {D3};
    \node[] at (1.6*3+.6,1.6*3+.6) {$ \cdot^{\cdot^{\cdot}}$};
\draw[->] (-2.2,.4) -- (-2.2+.9,.4) node[anchor= north] {$x$};
\draw[->] (-2.2,.4) -- (-2.2,1.3) node[anchor= east] {$y$};
\end{tikzpicture}
    \caption{A cross-section of the branes in the $x$-$y$ plane. Both the $x$ and $y$ direction are periodic. The entire plane is wrapped by $N$ D3-branes, which are shaded blue. The D5$_\alpha$-branes appear as point-like defects embedded inside the D3s.}
    \label{fig:branelayoutD2D2p}
\end{figure}
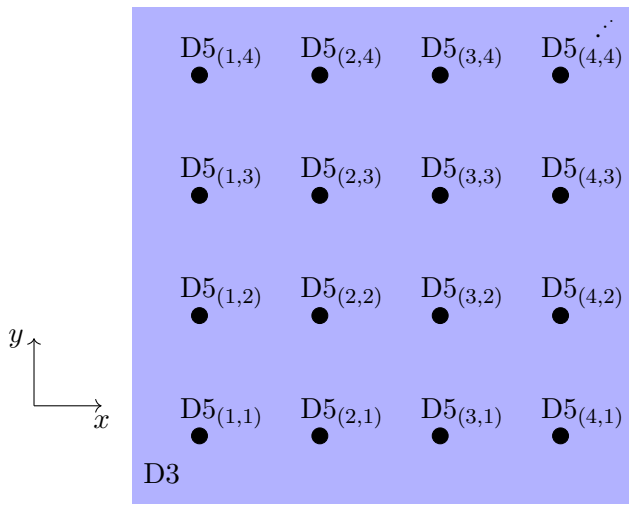

For our decoupling limit, we can start by taking $l_s\to 0$ (and $\epsilon \to 0$ in such a way that $M$ is fixed and small but non-zero). The world-volume fields on the D5$_\alpha$-branes will decouple because their (5+1)D Yang-Mills coupling goes like $g_{\mathrm{YM}}^2 \sim g_sl_s^2$. The $\textsl{U}(1)_\alpha$ gauge group on their world-volume then appears as a  \emph{global} symmetry from the perspective of the theory which governs the D3-D3 and D3-D5$_\alpha$ strings. The defect fermions $\chi_{\alpha,b}(t)$ have charge $1$ under $\textsl{U}(1)_\alpha$, and are neutral with respect to $\textsl{U}(1)_{\beta}$ when $\alpha\neq \beta$. We can therefore interpret the resulting $\prod_\alpha \textsl{U}(1)_\alpha$ global symmetry group as a point-like subsystem symmetry group once we promote $\alpha$ to a two-dimensional spatial coordinate. In order to keep the mass $M$ finite in this limit, we must then take $\epsilon\to 0$ simultaneously.
This setup is summarized in Table \ref{fig:D2D2p} and Figure \ref{fig:branelayoutD2D2p}.

The D3-D3 and D3-D5$_\alpha$ strings then furnish a Lagrangian of the same schematic form as Eq.\ \eqref{eqn:instantaneouscoulomb}. The difference is that the (3+1)D part will be described by (a non-relativistic limit of) $\mathcal{N}=4$ super Yang-Mills theory, while the quantum mechanical defect probes will be described by the fermionic fields $\chi_{\alpha,b}$ and are coupled in a way which preserves some supersymmetry. The decoupling limit can then be taken as in our non-supersymmetric prototype; we leave the determination of the precise supersymmetric Lagrangian to future work.

\section{Infinite-component Chern--Simons theories}\label{sec:ics}

The constructions of the previous sections were all supersymmetric and gapless. Supersymmetry is somewhat unnatural from the perspective of condensed matter physics, and it is also desirable to be able to produce gapped models. In this final section, we offer a brane construction which, although supersymmetric, is gapped and flows to a topological Abelian Chern-Simons theory in the IR (so that SUSY is trivially realized). We will see that, by taking the number of gauge fields to be large, we will be able to interpret this model as a theory of fractons in one dimension higher, following the treatment of infinite-component Chern--Simons theories in \cite{shirley2020twisted,ma2020fractonic}.  

To realize these models on branes, we rely on the well-established engineering of (2+1)D $\mathcal{N}=2$ supersymmetric quiver Chern-Simons gauge theories from ``brane tilings''  \cite{hanany1998realization,hanany1998brane,aganagic2000mirror, feng2002toric, franco2006brane, hanany2008tilings, Hanany:2008gx, hanany2009brane,imamura2008quiver}; we will argue that such brane configurations can be gapped and described at low energies by Chern-Simons theories with large and non-trivial K-matrices. Our main example is closely related to (an orbifold of) the ABJM theory, which was one of the first examples of AdS$_4$/CFT$_3$ holography coming from string/M-theory \cite{Aharony:2008ug,aharony2008fractional,benna2008superconformal}. 

The content of this section has some similarities with \cite{Razamat:2021jkx}. See also \cite{cho2020m} for work on engineering gapped topological phases from branes in M-theory. 

\subsection{Field theory}

Part of the inspiration for considering iCS theories is the success of ordinary Abelian Chern--Simons theory in describing Abelian topological order in (2+1)D \cite{wen1990topological,wen1992classification}. The data which specifies the latter kind of theory is a symmetric, integer $L\times L$ matrix $K_{IJ}$ which defines the Lagrangian as
\begin{align}\label{eqn:AbelianCS}
    \mathcal{L}_K = \frac{1}{4\pi}\sum_{I,J=1}^L K_{IJ} \epsilon^{\mu\nu\rho}a^I_\mu \partial_\nu a_\rho^J
\end{align}
where the $a^I$ are (2+1)D $\textsl{U}(1)$ gauge fields, with $I,J=1,\dots,L$.  
One of the reasons that this is a useful formalism is that the various universal properties of the topological phase --- including ground state degeneracies, fusion, braiding, edge states, and so on --- can be cleanly extracted just from $K_{IJ}$. 

The philosophy of iCS theory is to take $L$ to be large in Eq.\ \eqref{eqn:AbelianCS}, and to interpret $I$ as a (discrete) spatial coordinate of an emergent dimension. One can think of $I$ as labeling different (2+1)D layers which span this fourth dimension, and which are coupled together through the off-diagonal entries of the matrix $K_{IJ}$. The resulting theory then often gives rise to fracton order in (3+1)D. Indeed, consider the somewhat trivial example of a diagonal K-matrix
\begin{align}\label{eqn:diagonal}
    K = \left(\begin{array}{ccccc} \ddots & & & & \\ & k & & &  \\ & & k & &  \\ & & & k & \\ & & & & \ddots
    \end{array}\right).
\end{align}
On a $T^3$ spatial torus, this theory has a ground state degeneracy equal to 
\begin{align}
    \mathrm{GSD} = |\det K| = |k|^L
\end{align}
whose logarithm scales linearly with the size of the emergent dimension. The model also clearly admits planons:  these are simply the anyons of each constituent layer, which are able to move around within the layer to which they belong, but cannot hop from one layer to the next. We see that even in this very simple model, we recover two properties already that are characteristic of fracton order. 

To get a more interesting model, we can imagine coupling nearest neighbor layers by populating the entries of the K-matrix which are directly adjacent to the diagonal, 

\begin{align}\label{eqn:offdiagonal}
    \widetilde{K}(k) \equiv \left(\begin{array}{ccccc} k & 1 & & & 1 \\ 1 & k & 1 & &  \\ & \ddots & \ddots & \ddots &  \\ & & 1 & k & 1 \\ 1 & & & 1 & k
    \end{array}\right).
\end{align}
The reason for the non-zero entries in the top-right and bottom-left corners is that we take the emergent dimension to be periodic. All other entries are zero. 

The behavior of the theory depends sensitively on the choice of $k$. For example, when $k=3$, the theory is gapped, while when $k=2$, the theory is gapless. Interestingly, the model with $k=3$ was studied in \cite{qiu1989phases,qiu1990phase,naud2000fractional,naud2001notes} in the context of coupled fractional quantum Hall layers. From the perspective of fracton physics, it gives an example of \emph{non-foliated} fracton order. This may seem surprising given the clear presence of (2+1)D layers in the description of the system, however the word ``foliated'' here means something precise \cite{shirley2018fracton}. Specifically, one says that a (3+1)D theory is foliated in a particular direction if the theory with height $\Delta$ in that direction is local-unitary equivalent to the same theory with height $\Delta-\delta$, plus a decoupled block of size $\delta$. In the context of iCS theories, ``local unitary equivalence'' is implemented by a mapping $K\to WKW^T$ where $W$ is a finite-depth $\textsl{GL}(L,\mathbb{Z})$ matrix; that is, an integer matrix which can be decomposed into a finite product of local, block diagonal, integer, general linear matrices. Under this definition of ``foliated'', it is not obvious that every iCS theory should be foliated, and indeed there is no finite-depth $W$ which is able to decouple a layer in this manner when $k=3$. See \cite{ma2020fractonic} for further details on this model.

Achieving the exact K-matrix from Eq.\ \eqref{eqn:offdiagonal} in a brane setup is subtle for reasons that we will explain; however we will be able to realize a large class of K-matrices with similar physics. For concreteness, an example we will focus on is
\begin{align}\label{eqn:KmatrixBrane}
    K(k) = \left(
    \begin{array}{rrrrrrrrrrrr}
    \ddots & \\
    &a & 2 & & -1 & & & \\
    &2 & b & 1 & \\
    && 1 & a & 2 & & -1 \\
    &-1 &  & 2 & b & 1 & \\
    &&&& 1 & a & 2 & & -1 \\
    &&& -1 & & 2 & b & 1 \\
    &&&&& &1 & a & 2\\
    &&&&&-1 & &2 & b \\
    &&&&&&&&&\ddots
    \end{array}
    \right)
\end{align}
where $a = k-2$ and $b=-k-2$. It is straightforward to check, using the techniques of \cite{ma2020fractonic}, that this K-matrix defines another example of non-foliated fracton order.

One of the questions which was addressed in \cite{ma2020fractonic} was the extent to which theories like $\mathcal{L}_{K(k)}$ and $\mathcal{L}_{\widetilde{K}(k)}$ are truly local in the emergent dimension. This question was answered in the affirmative by constructing a local lattice model which goes over to $\mathcal{L}_K$ in the continuum limit, for any quasi-diagonal\footnote{A family of K-matrices whose size is tending to infinity is quasi-diagonal if every non-zero element is within some fixed distance from the diagonal.} choice of $K$ with bounded entries. In the next section, we will see this same fact from another perspective: $\mathcal{L}_{K(k)}$ is a local (3+1)D theory because it arises in a local way on D-branes. 

Before moving on, it will be useful to understand how such Chern-Simons theories can arise as the IR fixed points of certain (2+1)D supersymmetric field theories. The supersymmetric field theories in question are $\mathcal{N}=2$ quiver Chern-Simons theories, which we will take throughout to be Abelian. Every theory we consider in the rest of this paper will consist of the following ingredients. First, there will be $N$ Abelian vector-multiplets, each consisting of a $\textsl{U}(1)$ gauge field $a^I$ and its superpartners, which are a Dirac fermion and a real scalar. We will give the gauge fields in these multiplets Chern-Simons levels $k^I$; assuming these are all non-zero, the photon and its superpartners all gain a mass, so that all that is left of the vector-multiplet at low energies is simply the gauge field $a^I$ governed by a Chern-Simons term at level $k^I$. In other words, the IR is described by a CS theory with a diagonal K-matrix. 

In order to obtain off-diagonal elements in the K-matrix, we can add matter chiral multiplets, each of which consists of a complex scalar and a Dirac fermion. We take each matter chiral multiplet to transform in a bifundamental representation of a pair of gauge fields (i.e.\ we take each to have charge $+1$ with respect to some $a^I$ and charge $-1$ with respect to some $a^J$). If we give these chiral multiplets a real mass and integrate them out, then they produce contributions to the K-matrix \cite{redlich1984gauge,redlich1984parity}. Depending on the \emph{sign} of their mass, they either contribute $+1$ to $K_{II}$ and $K_{JJ}$ while contributing $-1$ to $K_{IJ}$ and $K_{JI}$; or they contribute $-1$ to $K_{II}$ and $K_{JJ}$ while contributing $+1$ to $K_{IJ}$ and $K_{JI}$. We note that to obtain off-diagonal K-matrix entries in this manner, it is crucial that the theory have $\mathcal{N}=2$ supersymmetry as opposed to, say, $\mathcal{N}=4$ supersymmetry. In a theory with $\mathcal{N}=4$ supersymmetry, the matter fields live in hypermultiplets, which in $\mathcal{N}=2$ language can be thought of as two chiral multiplets with charges and real masses of opposite sign. Thus, when a massive hypermultiplet is integrated out, it does not lead to any Chern-Simons terms. For review and discussion of related ideas, see e.g.\ \cite{dunne1999aspects,tong2000dynamics}.

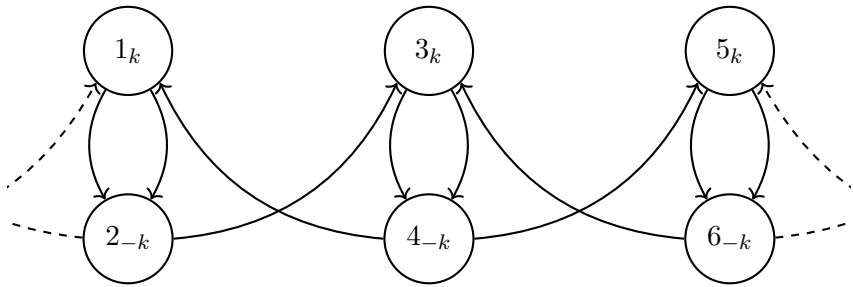
\begin{figure}
    \centering
    \begin{tikzpicture}
    \tikzset{vertex/.style = {shape=circle,draw,minimum size=.5cm, text width = 2em,align=center,thick}}
\tikzset{edge/.style = {->,thick}}
\tikzset{dotted/.style = {->,thick,dashed}}
\begin{scope}
    \clip (-1.6,-3.25) rectangle (9.6,.75);
    \node[] (0) at (-4,0) {};
    \node[] (-1) at (-4,-2.5) {};
    \node[vertex] (1) at (0,0) {$1_k$};
    \node[vertex] (2) at (0,-2.5) {$2_{-k}$};
    \node[vertex] (3) at (4,0) {$3_k$};
    \node[vertex] (4) at (4,-2.5) {$4_{-k}$};
    \node[vertex] (5) at (8,0) {$5_k$};
    \node[vertex] (6) at (8,-2.5) {$6_{-k}$};
    \node[] (7) at (12,0) {};
    \node[] (8) at (12,-2.5) {};
    \draw[edge, bend right] (1) to  (2);
    \draw[edge, bend left] (1) to  (2);
    \draw[edge, bend right] (3) to  (4);
    \draw[edge, bend left] (3) to  (4);  
    \draw[edge, bend right] (5) to  (6);
    \draw[edge, bend left] (5) to  (6);
    \draw[edge, bend right] (2) to (3.225);
    \draw[edge, bend left] (4) to (1.315);
    \draw[edge, bend right] (4) to (5.225);
    \draw[edge, bend left] (6) to (3.315);
    \draw[dotted, bend left] (2) to (0.315);
    \draw[dotted, bend right] (-1) to (1.225);
    \draw[dotted, bend right] (6) to (7.225);
    \draw[dotted, bend left] (8) to (5.315);
\end{scope}
    \end{tikzpicture}
    \caption{A quiver diagram which summarizes the field content of a (2+1)D $\mathcal{N}=2$ Chern-Simons gauge theory. Each node corresponds to a $\textsl{U}(1)$ vector multiplet, and each arrow corresponds to a bifundamental chiral multiplet. A node with a label $I_k$ means that $I$ is the index and $k$ is the level. The gauge fields alternate between having Chern-Simons terms at level $k$ and $-k$, and the bifundamental between node $I$ and $J$ has a positive real mass if $J>I$ and a negative real mass if $J<I$. In the IR, this theory is described by an Abelian Chern-Simons theory with K-matrix $K(k)$, Eq.\ \eqref{eqn:KmatrixBrane}.}
    \label{fig:ABJMorbifoldquiver}
\end{figure}

The field content described so far --- gauge fields in vector multiplets and matter fields in bifundamental chiral multiplets --- can be summarized by a quiver diagram. Each node of the quiver represents a vector-multiplet, and each arrow from the $I$th vector-multiplet to the $J$th vector-multiplet represents a bifundamental chiral with charge $+1$ with respect to $a^I$ and charge $-1$ under $a^J$. In principle, to completely determine the theory, one must also specify a superpotential, however for our purposes, the IR physics is sensitive only to the pattern of real masses, which we will specify separately. As an example, we note that the K-matrix $K(k)$ from Eq.\ \eqref{eqn:KmatrixBrane} arises from the quiver in Figure \ref{fig:ABJMorbifoldquiver}. This theory can be thought of as a massive deformation of a particular orbifold of the ABJM theory which appears in \S 6 of \cite{benna2008superconformal}. In the next section we will describe how it can be realized on branes.

\subsection{Type IIA brane construction}

We now turn to a brane realization of the particular iCS theory specified by the K-matrix in Eq.\ \eqref{eqn:KmatrixBrane}. In actuality, we will realize the supersymmetric quiver gauge theory (partially) specified by Figure \ref{fig:ABJMorbifoldquiver}, which flows in the IR to the desired iCS theory. One common way to realize quiver gauge theories is to work with branes in the presence of a transverse orbifold \cite{Douglas:1996sw}; however, for our purposes, it will be more convenient to work in a different duality frame where such configurations can be described by brane tilings.\footnote{In fact, the brane tilings should best be thought of as a graphic depiction of the matter content of this orbifold rather than a literal brane configuration, as only the former is under perturbative control within string theory.}

We work in Type IIA string theory and wrap the $x^3$ direction into a circle, and the $x^4$-$x^6$ plane into a torus $T^2$. A brane tiling is a D4/NS5 configuration. One starts by placing an NS5-brane in spacetime so that its world-volume spans the $x^0,\dots,x^3$ directions and wraps a surface $\Sigma$ in the $\mathbb{R}^2\times T^2$ coordinatized by $x^4,\dots, x^7$. This surface $\Sigma$ intersects the $x^4$-$x^6$ plane in a bipartite graph $G$, i.e.\ a graph whose vertices are all colored either black or white, and whose edges only connect vertices of different colors. One thinks of this graph as a tiling of the $x^4$-$x^6$ plane.  For each face $I$ of the graph/tiling $G$, one puts down a D4$^I$-brane which spans the $x^0,x^1,x^2$ directions and is strung inside the face $I$ in the $x^4$-$x^6$ plane. One then thinks of the world-volume theory on the D4$^I$-branes as a (2+1)D theory by taking the $T^2$ to be small. The brane configuration preserves 4 supercharges, and so the (2+1)D world-volume theory will have $\mathcal{N}=2$ supersymmetry.

\begin{table}
\begin{center}
\begin{tabular}{c|ccc|ccccccc}
&$t$&$x$&$y$&$3$&$4$&$5$&$6$&$7$&$8$&$9$ \cr
\hline
D4$^{I}$ & \textbf{x}&\textbf{x}&\textbf{x}& $\epsilon^I$  &$|\textbf{x}|$&&$|\textbf{x}|$& && \\
NS5 & \textbf{x}&\textbf{x}& \textbf{x} &\textbf{x}& \multicolumn{4}{c}{-------- $\Sigma$ --------} \\
\end{tabular}
\caption{A diagram which encodes the layout of branes in spacetime. The $x^3$ direction is a circle, and $x^4$, $x^6$ coordinatize a torus $T^2$. The indices run from $I=1,\dots, L$. An \textbf{x} denotes a direction that the brane spans. An $|\textbf{x}|$ denotes a direction that the D4$^I$-brane spans but is bounded on both sides by the NS5-brane. The D4$^I$-branes are located at the positions $\epsilon^I$ in the $x^3$ direction, and $\Sigma$ denotes a surface embedded in the $x^4,\dots,x^7$ directions that is wrapped by the NS5-brane.  }\label{table:branetilings}
\end{center}
\end{table}

A quiver can be naturally read off from a brane tiling as follows. First, strings stretching from a D4$^I$-brane to itself contribute a vector-multiplet worth of field content; thus, nodes of the quiver are in correspondence with faces of the graph $G$. The D4$^I$-brane action receives a contribution from a term of the form
\begin{align}
    S_{\mathrm{D}4^I} \supset \frac{1}{2\pi} \int_{\partial \mathrm{D}4^I} a^I\wedge da^I \wedge d\phi
\end{align}
where $\phi$ is the compact scalar living on the NS5-brane which corresponds to the embedding in the $x^{11}$-direction in the M-theory picture. Thus, by giving this scalar a non-trivial profile, one can induce Chern-Simons levels $k^I$ for the gauge fields in the vector-multiplets \cite{imamura2008quiver}. The levels so-obtained are subject to the constraint that 
\begin{align}
    \sum_I k^I = 0
\end{align}
which is why it is more challenging to produce a K-matrix like the one in Eq.\ \eqref{eqn:offdiagonal}. It may be possible to relax this condition by working in massive Type IIA (i.e.\ by turning on a Romans mass/zero-form flux \cite{romans1986massive,gaiotto2010gauge,aharony2010massive}), or relatedly by incorporating D8-branes \cite{hanany1998chiral,fujita2009fractional,brunner1998branes}, however it appears difficult to accomplish this without breaking supersymmetry or introducing extra unwanted gapless matter. We leave this generalization as an interesting direction for future work.

The arrows of the quiver diagram are also encoded in the brane tiling. Whenever two faces $I$ and $J$ are adjacent, i.e.\ separated by an edge of the graph $G$, there is an arrow from $I\to J$ or $J\to I$. The edge which separates $I$ and $J$ has two vertices as its endpoints, one black and one white, and the direction of the arrow which crosses this edge is such that the black vertex is on the left. The bifundamental chiral multiplet corresponding to this arrow encodes the physics of strings stretching across the NS5-brane from D4$^I$ to D4$^J$. One can give this bifundamental chiral multiplet a real mass by ensuring that D4$^I$ and D4$^J$ are separated in the $x^3$ direction, so that strings which stretch from D4$^I$ to D4$^J$ must stretch across a non-zero distance. The sign of this real mass depends on whether the location in the $x^3$ direction is increasing or not in the direction of the arrow. In principle, the full superpotential of the world-volume theory is encoded in the brane tiling as well, however we will not need this for our purposes.

\begin{figure}
    \centering
\begin{tikzpicture}
\tikzset{edge/.style = {->,thick}}
\fill[blue!30] (0,-1.6) rectangle (1.6*5,3.2);
\draw[step=1.6cm] (0,-1.61) grid (1.6*5,3.21);
\node[] (L) at (1.6/2,1.6/2) {D5$^{L}$};
\node[] (Lm1) at (1.6/2,-1.6/2) {D5$^{L-1}$};
\node[] (Lm1p) at (1.6/2,1.6+1.6/2) {D5$^{L-1}$};
\node[] (1) at (1.6/2+1.6,1.6/2) {D5$^{1}$};
\node[] (2) at (1.6/2+1.6,-1.6/2) {D5$^{2}$};
\node[] (2p) at (1.6/2+1.6,1.6+1.6/2) {D5$^{2}$};
\node[] (4) at (1.6/2+1.6*2,1.6/2) {D5$^{4}$};
\node[] (3) at (1.6/2+1.6*2,-1.6/2) {D5$^{3}$};
\node[] (3p) at (1.6/2+1.6*2,1.6+1.6/2) {D5$^{3}$};
\node[] (5) at (1.6/2+1.6*3,1.6/2) {D5$^{5}$};
\node[] (6) at (1.6/2+1.6*3,-1.6/2) {D5$^{6}$};
\node[] (6p) at (1.6/2+1.6*3,1.6+1.6/2) {D5$^{6}$};
\node[] at (1.6/2+1.6*4,1.6/2) {$\cdots$};
\node[] at (1.6/2+1.6*4,1.6+1.6/2) {$\cdots$};
\node[] at (1.6/2+1.6*4,-1.6/2) {$\cdots$};
\draw[->] (-2.2,.4) -- (-2.2+.9,.4) node[anchor= north] {$x^6$};
\draw[->] (-2.2,.4) -- (-2.2,1.3) node[anchor= east] {$x^4$};
\draw[edge] (L) to (1);
\draw[edge] (Lm1p) to (L);
\draw[edge] (Lm1) to (L);
\draw[edge] (2) to (Lm1);
\draw[edge] (2p) to (Lm1p);
\draw[edge] (1) to (2);
\draw[edge] (1) to (2p);
\draw[edge] (2) to (3);
\draw[edge] (2p) to (3p);
\draw[edge] (6p) to (3p);
\draw[edge] (3) to (4);
\draw[edge] (3p) to (4);
\draw[edge] (4) to (1);
\draw[edge] (6) to (3);
\draw[edge] (4) to (5);
\draw[edge] (5) to (6);
\draw[edge] (5) to (6p);
\foreach \x in {0,...,2}
\foreach \y in {0,...,1}
    {\filldraw (\x*3.2,\y*3.2) circle (3pt); 
    \filldraw (\x*3.2+1.6,\y*3.2-1.6) circle (3pt);
    \filldraw[fill=white] (\x*3.2,\y*3.2-1.6) circle (3pt);
    \filldraw[fill=white] (\x*3.2+1.6,\y*3.2) circle (3pt);
    }
\end{tikzpicture}
    \caption{A visualization of a brane tiling in the $x^4$-$x^6$ plane. The D5$^{I}$-branes form boxes which tile the plane and are separated by the NS5-brane. The grid has period 2 in the $x^4$ direction and period $L/2$ in the $x^6$ direction.}
    \label{fig:braneboxes2}
\end{figure}
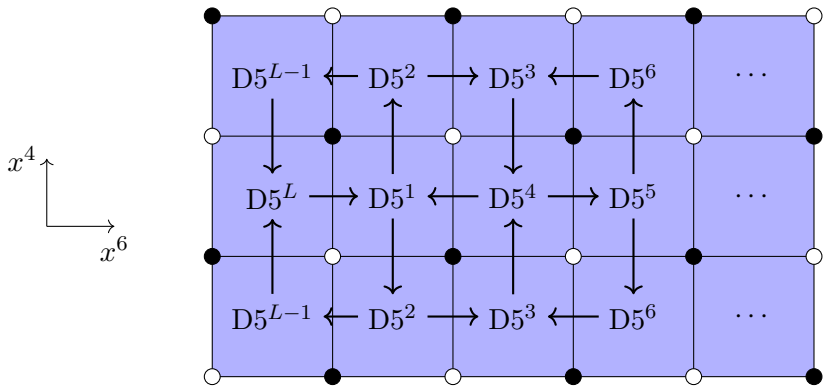

Given these rules, it is clear that brane tilings can produce a rather large class of iCS theories with quasi-diagonal K-matrices. Consider for example the simple tiling in Figure \ref{fig:braneboxes2}. We take the $T^2$ to have a large width but a small height, and tile it with a $2\times \frac{L}{2}$ grid of boxes. It is clear that the quiver associated to this tiling is none other than the ABJM orbifold given in Figure \ref{fig:ABJMorbifoldquiver}. To give the bifundamental chirals a real mass, we place the D5$^I$-brane at a position $\epsilon^I = \frac{2\pi R}{L}I$ in the $x^3$-direction, where $R$ is the radius of the circle in that direction. All in all, we end up engineering an iCS theory with K-matrix given in Eq.\ \eqref{eqn:KmatrixBrane}.

\section{Conclusion}\label{sec:conclusion}

In this paper, we have highlighted how D-branes are related to three different kinds of continuum methods which have recently appeared in the literature in connection with fractons and exotic symmetries. There are a number of remaining questions for future research.  
\begin{enumerate}
    \item It is known that there are lattice models which can be described equally well in the continuum by both foliated field theories and exotic field theories; for example, the X--cube model admits both kinds of presentations. It would be desirable to establish the equivalence between the foliated field theory approach and the exotic field theory approach directly in the continuum. We believe some of the methods used in \S\ref{subsec:oneformnonsusy} could be put to good use in this pursuit. 
    
    Related to this, it would be interesting if some of the $\textsl{U}(1)$ foliated field theories discussed in this paper could be given a presentation more along the lines of \cite{seibergshao1,seibergshao2,seibergshao3}. For example, it is tempting to speculate that there is a relationship between the model in \S\ref{subsubsec:linearfoliatednonsusy} and the XY-plaquette model which was studied in \cite{seibergshao1}, due to the similarity in the structure of their symmetries and dualities. The $\mathbb{Z}_N$ version of this statement is in fact true \cite{brdw}. 
     \item Brane constructions are often a useful starting point for studying field theories using holographic duality. Many of the models in this paper are variations on brane configurations with known holographic duals. For example, the D3/D5 system which was used as an ingredient in \S\ref{subsec:planar} admits a holographic description \cite{karch2001localized,Karch:2000gx} in terms of an AdS$_4$ brane inside an AdS$_5$ spacetime \cite{Karch:2000gx}, generalizing the original holographic duality between the world-volume of D3-branes and AdS$_5$ gravity \cite{maldacena1999large}. 
     As another example, models closely related to the one studied in \S\ref{subsec:pointlikesusy} were investigated holographically in \cite{kachru2011adventures,kachru2010holographic}. It would be interesting to revisit these constructions with fractons and exotic symmetries in mind. We note previous attempts to relate fractons to gravity and holography \cite{pretko2017emergent,yan2019hyperbolic,yan2019hyperbolic2,yan2020geodesic}. 
     \item Throughout most of the paper, we specialized to $\textsl{U}(1)$ symmetries. It should be possible to incorporate $\textsl{U}(N)$ groups by increasing the number of coincident branes in various places in our constructions. Can $\textsl{SO}(N)$ and $\textsl{Sp}(N)$ groups be accommodated by incorporating orientifold planes?
     \item The various continuum approaches feature several interesting phenomena, including discontinuous field configurations, exotic symmetries, duality, particles with restricted mobility, and UV/IR mixing. It would be interesting to study in further depth how these properties are imprinted in the brane physics.
    \item It should be possible to modify the brane construction of \S\ref{sec:eft} to produce theories with linear and planar subsystem symmetries. This also suggests that there are supersymmetric extensions of some of the models considered in \cite{seiberg0,seibergshao1,seibergshao2,seibergshao3}, though we have not attempted to write their actions down. Some supersymmetric exotic field theories are known \cite{3+1dsusyfractons}, but these theories look rather different from the brane systems we consider. It would be interesting to work out these supersymmetric effective actions more systematically, and compute observables which are protected by supersymmetry, like BPS indices.
    \item In this paper, we have mostly worked with 0-form global (subsystem) symmetries and 1-form gauge fields. The descriptions of many fracton theories (like the foliated field theory description of the X-cube model) involve generalized global symmetries and higher-form gauge fields as ingredients. Given the abundance of e.g.\ higher-form gauge fields in string theory, it is tempting to try to find string embeddings of such fracton theories.
    \item In \S\ref{subsec:planar}, we were able to relate the duality properties of certain foliated field theories to S-duality of Type IIB string theory. Are there more examples? What aspects of the non-supersymmetric duality web in (2+1)D \cite{seiberg2016duality,Karch:2016sxi} can be generalized to theories with planar $\textsl{U}(1)$ subsystem symmetries in (3+1)D?
\end{enumerate}

\section*{Acknowledgements}
We thank Milind Shyani for participation in the initial stages of this project. We gratefully acknowledge insightful exchanges with John McGreevy, Daniel Ranard, Jonathan Sorce, Dominic Williamson, and Max Zimet, and thank the members of the Simons Collaboration on Ultra-Quantum Matter for inspiring conversations and seminars. We especially thank John McGreevy for crucial comments on an earlier draft. BR thanks Dominic Williamson for collaborations which have strongly shaped how he thinks about the subject. HG is very grateful to his parents and recommenders.

\paragraph{Funding information}
 The work of AK was supported, in part, by the U.S.~Department of Energy under Grant No.~DE-SC0011637 and by a grant from the Simons Foundation (Grant 651440, AK). The work of SK was supported by the Simons Collaboration on
Ultra-Quantum Matter, which is a grant from the Simons Foundation (651440, SK).
It was also supported by the Department of Energy under grant DE-SC0020007, and a Simons Investigator Award. The work of BR was supported by NSF grant PHY 1720397. RN is funded by NSF Fellowship DGE-1656518, Department of Energy under grant DE-SC0020007, and an EDGE grant from Stanford University.

\bibliography{arxiv_v2.bib}

\nolinenumbers

\end{document}